\documentclass[aps,prd,preprint,superscriptaddress,showkeys,nofootinbib,floatfix]{revtex4-2}
\usepackage{graphicx} 

\usepackage{amsmath,amssymb,amsfonts}
\usepackage{bm}
\usepackage{boxedminipage}
\usepackage{enumitem}
\usepackage{float}
\usepackage{physics}
\usepackage{hyperref}
\usepackage{colonequals}
\usepackage{xcolor}
\usepackage{textcomp,gensymb}
\usepackage{capt-of} 

\newcommand*{\logeq}{\Leftrightarrow}
\def\rbm#1{\xrbm#1\relax^\relax\valign}
\def\xrbm#1^#2\relax#3\valign{%
\mathbf{#1}\ifx\valign#2\valign\else^{\mathbf{#2}}\fi}

\def\xrbu#1^#2\relax#3\valign{%
\hat{\rbm{#1}}\ifx\valign#2\valign\else^{\rbm{#2}}\fi}

\makeatletter
\let\orig@appendix\appendix
\renewcommand{\appendix}{\def\p@section{\appendixname~}\def\p@subsection{\appendixname~\thesection.}\orig@appendix}
\makeatother

\newcommand{\Xmax}{X_{\max}}
\newcommand{\Xmaxpos}{\rbm{X_{\max}}}

\begin{document}
\title{Amplitude informed interferometry: Reconstruction of air shower direction and maximum with low signal-to-noise radio transients}

\author{A. Ferrière}
\affiliation{Sorbonne Université, CNRS, Laboratoire de Physique Nucléaire et des Hautes Energies (LPNHE), 4 Pl. Jussieu, 75005 Paris, France}
\affiliation{Université Paris-Saclay, CEA-LIST, F-91120 Palaiseau, France}

\author{V. Decoene}
\affiliation{SUBATECH, IN2P3-CNRS, Nantes Université, IMT Atlantique, 4 rue Alfred Kastler, 44300 Nantes, France}

\author{S. Cabana Freire}
\affiliation{Departamento de F\'isica de Particulas \& Instituto Galego de F\'isica de Altas Enerx\'ias, Universidad de Santiago de Compostela, 15782 Santiago de Compostela, Spain}

\author{O. Martineau-Huynh}
\affiliation{Sorbonne Université, CNRS, Laboratoire de Physique Nucléaire et des Hautes Energies (LPNHE), 4 Pl. Jussieu, 75005 Paris, France}
\affiliation{Sorbonne Universit\'{e}, UPMC Univ. Paris 6 et CNRS, UMR 7095, Institut d'Astrophysique de Paris, 75014 Paris, France}

\date{\today}

\begin{abstract}
Reconstructing neutrino-induced air showers with sparse radio arrays is limited by low signal-to-noise ratios (SNR). In HERON, a next-generation instrument for Earth-skimming ultra-high-energy neutrinos, phased stations trigger standalone antennas on which the SNR is typically below 1, so per-antenna arrival times and amplitudes, the inputs of conventional methods, cannot be measured reliably. Interferometry recovers such weak signals through coherence across the array, but uses timing information alone. We show that this biases the reconstruction: for an array subtending a small angle at the source, the reconstructed direction follows the line of sight rather than the shower axis, limiting the angular resolution to $2.0^\circ$ for HERON-like events. We introduce an interferometric method that also exploits the expected signal footprint, weighting each antenna by the amplitude predicted by the Angular Distribution Function (ADF) model of the radio emission. We show analytically that these weights maximize the coherent SNR, and that the maximum lies at the true direction and shower-maximum position, which removes the bias. On simulated $\tau$-induced air showers in the HERON geometry with realistic Gaussian noise near trigger threshold, the method achieves median errors of $0.23^\circ$ on the arrival direction, nearly an order of magnitude better than timing-only interferometry, and $0.54$~km on the three-dimensional position of the shower maximum. A first fluence-based estimate of the shower energy yields a $37\%$ resolution. The method also sets a design requirement: inter-antenna time synchronization must be better than $3$~ns for sub-degree angular resolution, which enables neutrino astronomy. Furthermore, a $1$~ns synchronisation is needed for kilometer-level localization of the shower maximum.
\end{abstract}

\keywords{Ultra-high-energy neutrino; Air showers; Radio detection; Reconstruction; Interferometry}

\maketitle



\section{Introduction}
Ultra-high-energy (UHE) neutrinos, with energies above $10^{16}$\,eV, can be produced through hadronic processes in cosmic accelerators (see e.g.~\cite{snowmass_nu} for a review). They then travel unimpeded across cosmological distances while still pointing back to their sources, which makes them a unique probe of the most extreme astrophysical processes.

Their very low expected flux, however, calls for instruments of very large effective area. A promising strategy uses the Earth itself as a target: a UHE neutrino undergoing a charged-current interaction in the Earth's crust produces a daughter lepton which, for tau flavor and Earth-skimming trajectories, may emerge into the atmosphere and decay there, initiating an upward-going extensive air shower (EAS)~\cite{Fargion:2000iz}. The electrons and positrons of the shower pancake drift under the influence of the Earth's magnetic field, and the resulting radio emission adds up coherently to reach detectable levels from a few tens to a few hundreds of MHz~\cite{Schroder:2016hrv}. Cheap, robust and easy to deploy, radio antennas are well suited to instrumenting the gigantic areas that a UHE neutrino detector requires.

Two complementary architectures have been proposed for this purpose. The BEACON project~\cite{BEACON} aims at interferometric detection: the signals of $N\sim\mathcal{O}(10$--$100)$ antennas arranged in a dense array (diameter $d\sim\mathcal{O}(100$\,m)) are summed at DAQ level with time delays chosen so that a wave propagating in a given direction adds up coherently. As the noise adds incoherently, the signal-to-noise ratio (SNR) is boosted by a factor $\sqrt{N}$, and synthesizing several beams towards the directions from which neutrino-induced EAS are expected to emerge (typically a valley at the foot of the array) yields a very sensitive instrument with a large effective area. The GRAND project~\cite{GRAND:2018iaj} instead proposes a network of $20$ sub-arrays of $10\,000$\,km$^2$, totaling $200\,000$ standalone antennas. The distinctive signatures of neutrino-induced EAS (in particular the deep position of the shower maximum and the elongated footprint at ground) should then allow both an efficient rejection of backgrounds (cosmic-ray-induced EAS, anthropogenic sources) and an angular resolution of $\mathcal{O}(0.1\degree)$~\cite{GUELFAND2025}.

The HERON project~\cite{Kotera2025} combines the sensitivity of the phased-array approach with the reconstruction capability of a standalone array. It is planned for installation in the San Juan province of Argentina, where a valley $30$--$100$\,km wide separates two parallel mountain ridges of about $2000$\,m altitude. This encased geometry increases the volume of Earth available for neutrino interactions and, at the same time, constrains the region in which a genuine $\tau$-induced shower can develop. The detector consists of $24$ phased stations of $24$ high-gain antennas each, deployed $3$\,km apart along $\sim70$\,km of one crest, each station being surrounded by $15$ standalone antennas spaced by typically $500$\,m. With fewer than a thousand antennas in total, this hybrid layout reaches a sensitivity comparable to that of much larger projects such as GRAND10k ($10\,000$ antennas) or BEACON-24, in particular below $100$\,PeV where the neutrino flux is expected to peak~\cite{Kotera2025}.

Two reconstructed quantities are critical to these science goals. The first is the angular resolution: HERON aims at identifying individual neutrino sources against a near-isotropic background, and the better the angular resolution, the tighter the association that can be made with a plausible astrophysical counterpart. The second is the three-dimensional position of the shower maximum, $\Xmaxpos$, and specifically its distance to the array.\footnote{Throughout this paper, $\Xmaxpos$ denotes the \emph{position} in space of the shower maximum, and not the atmospheric depth $\Xmax$ at which the shower reaches its maximum.} This distance separates the two populations HERON must distinguish: in simulation, a shower induced by a $\tau$ decay reaches its maximum typically $50$\,km from the array, and up to $100$\,km, constrained by the topography of the terrain. Cosmic-ray showers with inclinations larger than $87\degree$ reach theirs at least $250$\,km away. A reconstruction of $\Xmaxpos$ at the few-kilometer level is therefore sufficient to determine the nature of the primary.

Data acquisition follows a hierarchical, two-level trigger. Each phased station coherently sums its $N_{\rm phased}=24$ antennas along several digital beams, boosting the SNR by $\sqrt{N_{\rm phased}}$ and allowing the station to trigger on pulses too weak to cross the detection threshold of a single antenna. It then instructs the surrounding standalone antennas to record their buffered waveforms around the expected arrival time, whether or not they would have self-triggered on the pulse. This external triggering is what distinguishes HERON from self-triggered radio arrays, and it comes at a price. A self-triggered antenna requires a pulse of typically $5\,\sigma_{\rm noise}$~\cite{GRAND:HW}, with $\sigma_{\rm noise}$ the standard deviation of the baseline, so that every antenna in an event carries, by construction, a signal clearly identifiable above its local noise floor. The station threshold adopted in this study (Sec.~\ref{sec:simu}) corresponds instead to an amplitude of only $\sim\!0.8\,\sigma_{\rm noise}$ on an individual antenna. A HERON event therefore consists of a set of waveforms in which no single antenna necessarily shows an identifiable pulse, even though the signal is present as a coherent structure spread across the array. Reconstruction methods based on per-antenna arrival times and amplitudes~\cite{GUELFAND2025}, which presume a detectable pulse on each antenna, are not applicable in this regime.

Interferometry offers a way around this limitation. By coherently summing the signals of the whole array with the time delays expected from a candidate source position, it recovers a pulse that is invisible on any single antenna, at the price of a scan over the possible source positions. The technique is already exploited in air-shower experiments such as Auger~\cite{AugerInterf}, LOFAR~\cite{LOFARInterf} and, more recently, SKA-Low~\cite{SKAInterf}. It relies, however, on timing information alone, and we show in this article that this leads to a systematic bias for an array subtending a small angle at the source: the reconstructed direction follows the line of sight from the barycenter of antennas to the source location.

The main purpose of this article is to study how interferometric methods can be applied to the radio data collected with a sparse antenna array triggered by a phased array, down to per-antenna SNR below $1$, in order to reconstruct the properties of neutrino-induced EAS. We first present in Sec.~\ref{sec:simu} the simulation dataset on which our study is based. In Sec.~\ref{sec:limits} we establish the limits of standard interferometric reconstruction for extended radio detectors, identifying in particular the bias described above. Sec.~\ref{sec:adfbf} introduces the {\it ADF-interferometry} method, in which each antenna is weighted by the amplitude predicted by the Angular Distribution Function (ADF) model~\cite{GUELFAND2025}. Sec.~\ref{sec:perf} quantifies the performance of the method and the impact of noise and of a finite timing resolution, and presents a proof of concept for the energy reconstruction. Detailed derivations are collected in the appendices.

\section{Simulation dataset}\label{sec:simu}
We use for this study a simulation set built for the HERON standalone array. The layout and noise treatment are generic enough that the conclusions extend to other sparse arrays.

The simulation framework is based on the DANTON code~\cite{DANTON}, which provides an accurate sampling of the $\tau$ leptons generated by neutrino interactions in the Earth, including energy losses and scattering. The SRTM Digital Elevation Data is used to model the topography around the targeted site, a $100$\,km-long mountain range with crest height between $2000$ and $3000$\,m a.s.l., $1000$\,m above a flat valley in a remote area of the San Juan province. 

\begin{figure}[!ht]
    \centering
    \includegraphics[width=1\textwidth]{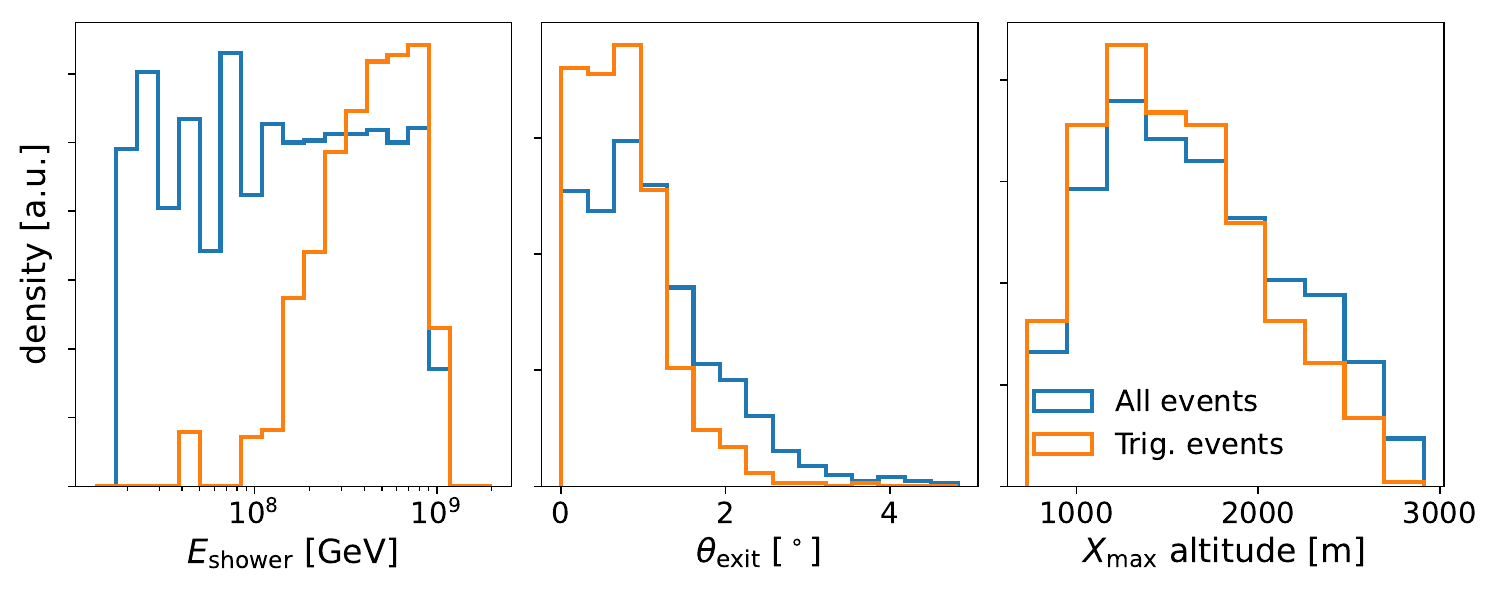}
    \caption{Energy ({\it left}), exit angle ({\it middle}) and decay altitude ({\it right}) distributions for the full simulation set (blue lines) and for the showers that trigger at least one phased station (orange lines).}\label{fig:sim_distribs}
\end{figure}

Resulting $\tau$ decays are drawn according to the branching fractions of the different decay modes, with arrival directions distributed randomly within a $3\degree$ half-angle cone from the array. A simulation set composed of $972$ $\tau$-neutrinos has been produced. Energies of the showers (non-neutrino decay product energies) range from $2\times10^{16}$~eV to $10^{18}$~eV (corresponding to $\nu$ energies from $10^{17}-10^{21}$~eV). Fig.~\ref{fig:sim_distribs} displays the distribution of energy of the $\tau$, exit angle and decay altitude.

\begin{figure}[!ht]
    \centering
    \includegraphics[width=0.75\textwidth]{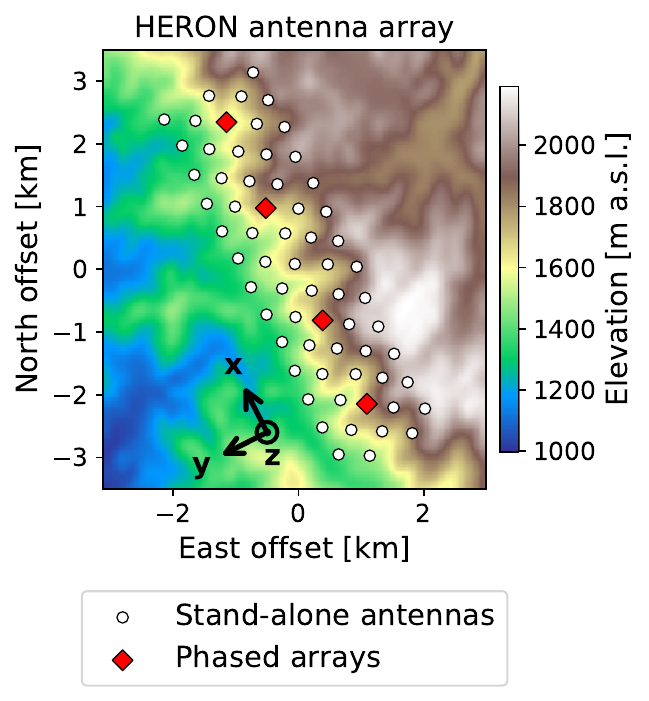}
    \caption{Layout of the HERON standalone array. The 61 standalone antennas are represented by white dots, while the 4 phased stations are represented by red squares. The color scale represents the elevation of the terrain in meters.}\label{fig_heron:layout}
\end{figure}

The decay products are then injected into the ZHAireS-RASPASS simulation code~\cite{alvarez2012,Tueros:2023mxa}, which simulates the development of the resulting EAS and computes the radio emission at 65 locations following the triangular layout (500,m spacing) shown in Fig.~\ref{fig_heron:layout}, and corresponding to a typical HERON standalone layout.
The simulated waveforms are computed along the three polarizations (South-North, West-East and Up) with a timestep of $0.1$~ns, giving a vector signal $\rbm{S}_i = (\,S^x_i(t), S^y_i(t), S^z_i(t)\,)$ for antenna $i$. The electric-field signal is then filtered to the $50$--$200$\,MHz band, the expected bandwidth of the HERON antennas, using a causal fifth-order Butterworth bandpass filter.

Gaussian noise can be added to the simulated waveforms, with a standard deviation $\sigma_{\rm noise} = 13~\mu$V\,m$^{-1}$, corresponding to the level expected from the Galactic background in this frequency range~\cite{Decoene2019}. As this is by far the dominant noise contribution~\cite{XuXing}, no other noise source is considered. Finally, Gaussian jitter with a standard deviation $\sigma_{\rm t} \in \{1, \ldots, 5\}$~ns can be added to the time tagging of each antenna, to emulate imperfect time tagging of the individual antenna measurements. 

All results presented in this paper are obtained on electric fields. The extension to voltages requires folding in the antenna response and the electronic chain, which would restrict the scope of the study to a specific antenna design. It is therefore left to a subsequent study. However, it is worth noting that~\cite{Guelfand:ICRC2025} shows that voltage footprints follow an amplitude distribution compatible with the ADF model we use here (see Sec.~\ref{sec:adftuning}). 

Two distinct SNR quantities are considered in this paper:
\begin{itemize}
    \item the \emph{observed} SNR, computed on the noisy waveforms, the only one accessible in real data. It is bounded from below, since the maximum of the noise is never zero;
    \item the \emph{true} SNR, computed on the noise-free waveforms, still assuming a $\sigma_{\rm noise} = 13~\mu$V\,m$^{-1}$ noise level. It can be lower than 1, meaning that the signal is buried in the noise.
\end{itemize}
Four of the 65 array positions defined in this study, represented by red squares in Fig.~\ref{fig_heron:layout}, stand for phased stations. For these we therefore compute the true SNR as $\sqrt{N_{\rm phased}}\times \mathrm{SNR}_{\rm single}$, with $N_{\rm phased}=24$ and $\mathrm{SNR}_{\rm single}$ the true SNR of the individual antenna.
We consider that a simulated event has triggered if at least one phased station has an SNR above $4$, corresponding to a per-antenna threshold of $\mathrm{SNR}_{\rm single}=0.82$, i.e. an amplitude of $10.6\,\mu$V\,m$^{-1}$. Among the $972$ events simulated in our data set, $548$ pass this trigger condition, as represented in Fig.~\ref{fig:sim_distribs}. The modest statistic used here is nevertheless sufficient to demonstrate the proof of principle of our method.

In the following sections the reconstruction methods are applied successively to the noise-free and to the noisy waveforms of these 548 events, and the results are compared to assess the performance in a realistic context. In the reconstruction presented below, all antennas (including the 4 phased stations) are treated in an identical way. This differs from realistic conditions in two ways.  First, one could expect that the 4 phased stations present a superior data quality compared to the 61 standalone units. Second, all 65 antennas are used rather than only those surrounding the triggering phased station(s).

\section{Limits of standard interferometric reconstruction}\label{sec:limits}

\subsection{Principle of interferometric reconstruction}
The interferometric reconstruction method for cosmic-ray air showers was introduced to reconstruct the particle profile and the direction of origin of air showers with high precision~\cite{HarmInterf,LOPESinterf}. It is inspired by radio astronomy, where the signals on multiple detectors are translated in time with specific delays to coherently add the radio emission of celestial bodies in the corresponding directions. Yet in the case of extended radio arrays the far-field approximation is not valid, as sources (the air showers) are closer than the Fraunhofer distance
\begin{align}\label{eq_heron:fraunhofer}
    d_F = \frac{2 L^2}{\lambda} \approx 20\,000~\text{km},
\end{align}
with $L\approx 6$~km the largest baseline of the array and $\lambda\approx 3.75$~m the wavelength of the signal (80~MHz). For a phased station of radius $L_{\rm phased}\approx0.1$~km, by contrast, $d_F\approx 5$~km: the far-field approximation does hold and classical beamforming remains valid for triggering at station level.

Rather than scanning the sky with a beam, the interferometric method for air showers reconstruction consists of scanning the possible source locations in the atmosphere. A coherent signal emitted at a candidate source location $\rbm{X_s}$ arrives at the antennas at different times, depending on the distance between the source and each antenna. The time taken to reach antenna $i$ at position $\rbm{p}_i$ from source position $\rbm{X_s}$ is
\begin{align}\label{eq_heron:delay}
    \Delta t_i(\rbm{X_s}) = \frac{n_i \|\rbm{X_s}-\rbm{p}_i\|}{c},
\end{align}
with $n_i$ the average refractive index between $\rbm{X_s}$ and $\rbm{p}_i$. Summing the signals received at each antenna with time delays given by Eq.~\eqref{eq_heron:delay} yields a coherent sum if a signal is indeed emitted at $\rbm{X_s}$. As the noise adds incoherently, the SNR then scales as $\sqrt{N}$. On the contrary summing signals from a point with no emission yields only noise. The summed signal $\rbm{S}$ is defined as
\begin{align}\label{eq_heron:standard_bf}
    \rbm{S} \left( \rbm{X_s};\ t \right) = \sum_i \rbm{S}_i\left( t + \Delta t_i(\rbm{X_s}) \right),
\end{align}
with $\rbm{S}_i$ the electric field signal on antenna $i$.

A metric quantifying the coherence of the summed signal, such as the maximum amplitude or the fluence of $\rbm{S} \left( \rbm{X_s},\,t \right)$, can then be computed as a function of $\rbm{X_s}$, yielding a 3D interferometric map. Since every point along the shower emits coherently, this map shows a coherent structure along the shower axis, from which the direction of arrival can be reconstructed by fitting a line through the points of maximum coherence.

To illustrate the limitations of standard interferometry in a HERON-like context, we use a proof-of-concept reconstruction adapted from~\cite{HarmInterf}.

\begin{figure}[!ht]
    \centering
    \includegraphics[width=0.98\textwidth,trim={0 6.0cm 0 6.0cm},clip]{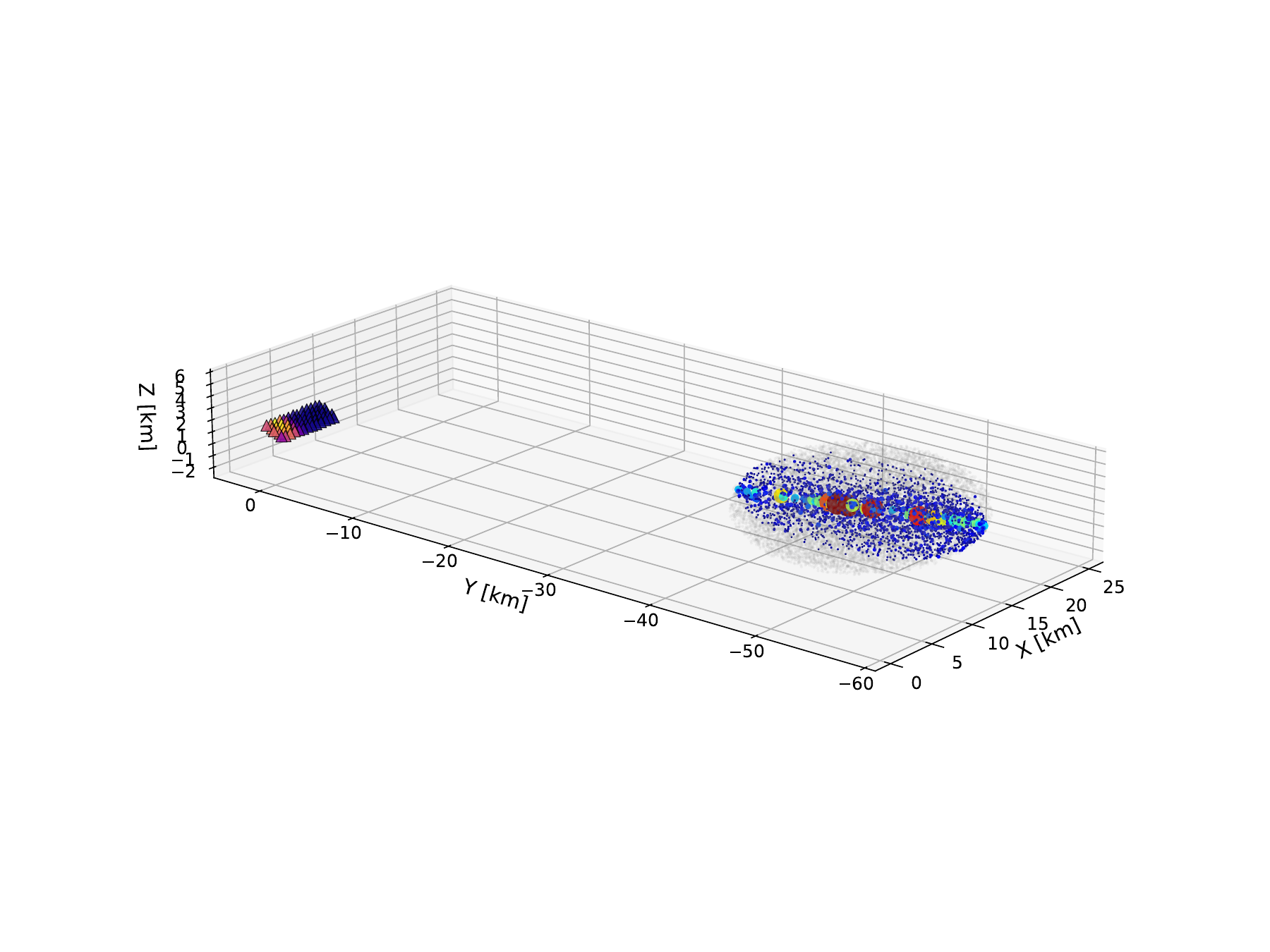}
    \caption{3D interferometric map of a $\tau$-induced air-shower. The color scale represents the coherence value. The points with the highest coherence are concentrated along the shower axis. Points passing the $2\%$ threshold are represented by colored dots, others by gray dots. A line is fitted to the points passing the $2\%$ selection to reconstruct the direction of arrival of the signal. The antennas forming the detector are represented by colored triangles.}
    \label{fig_heron:classical_interf_map}
\end{figure}

In the original work~\cite{HarmInterf}, the EAS trajectory was reconstructed by computing the interferometric map on three distinct horizontal planes, then fitting a straight line between the three points of maximum coherence. 
To work with near horizontal trajectories, we sample the 3D space randomly around the true shower-maximum position $\Xmaxpos$. This location is known from simulation and would be inferred from the triggered beam of the phased station. The search volume is an ellipsoid of $10$\,km radius in the horizontal plane and 2~km radius along the vertical axis (wide enough to contain all location with significant coherence). 
The main direction is then obtained by fitting a line minimizing distance to the $2\%$ of points with the highest coherence, each weighted by its coherence value. The $2\%$ threshold is low enough to exclude low-coherence points that would bias the fit, yet high enough to retain a sufficient number of points to constrain the line. This is illustrated in Fig.~\ref{fig_heron:classical_interf_map}.

\subsection{Limitations: the line of sight direction degeneracy}\label{sec:degeneracy}
The proposed method works well for cosmic-ray air showers, but has a major limitation: the reconstruction uncertainty along the line of sight is larger than the longitudinal elongation of the shower itself. In the following the line of sight direction is defined as the axis joining the source to a point $\rbm{p_b}$ in the array defined in~\ref{app:degeneracy} (and approximated by the average position of the antenna), hereafter the ``line of sight direction'', 

This can be quantified from a purely geometrical argument, derived in~\ref{app:degeneracy}. Writing $\delta \Delta t$ for the uncertainty on the difference of signal times between two antennas, the resulting uncertainty on the source position along the longitudinal axis is
\begin{align}\label{eq_heron:deltaD}
    \delta D \approx \frac{2c\,\delta \Delta t}{\sqrt N \omega_{\rm eff}^2},
\end{align}
with $\omega_{\rm eff}$ the effective opening angle between $\rbm{p_b}$ and the antennas, defined in~\ref{app:degeneracy}, seen from the source position $\rbm{X_s}$.
\begin{figure}[!ht]
    \centering
    \includegraphics[width=0.8\textwidth]{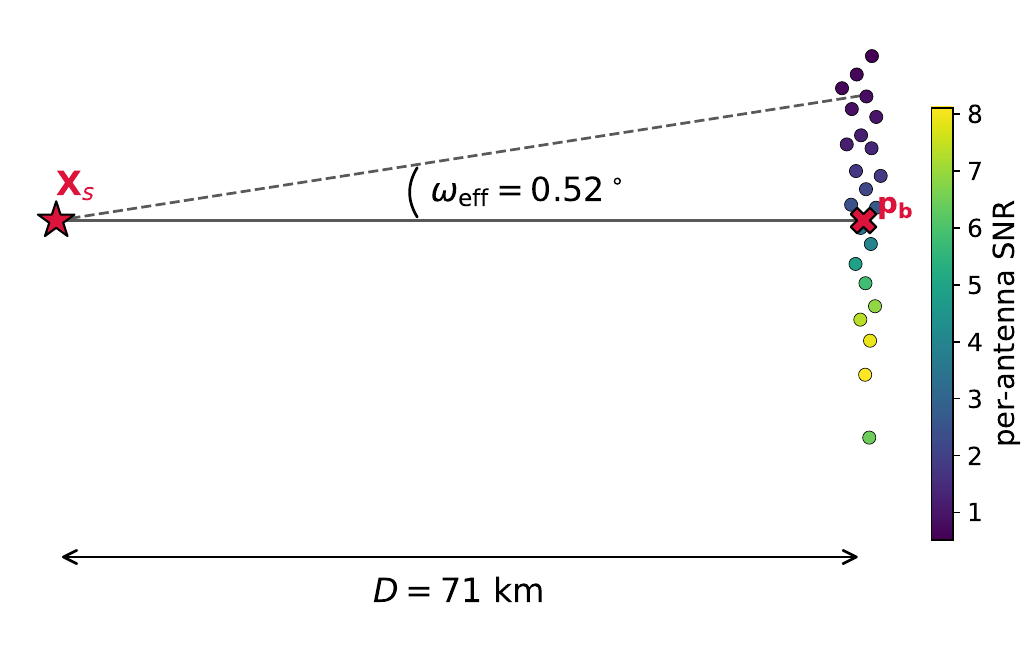}
    \caption{Example of an event with significant longitudinal bias. The effective opening angles $\omega_{\rm eff} = \left({\frac{1}{N} \sum_i \Delta \omega_i^4 }\right)^{1/4}$ between $\rbm{p_b}$ and each antenna $i$ seen from the source position $\rbm{X_s}$ is represented. See text for details.}
    \label{fig_heron:uncertainty_computation}
\end{figure}
Fig.~\ref{fig_heron:uncertainty_computation} illustrates this effect on a specific example. For this event, $23$ antennas have a signal with true SNR above $0.6$. 
Below this value, even the coherent sum of all $65$ antennas would not exceed an SNR of $0.6\sqrt{65}\approx 5$, required to detect the summed signal and perform reconstruction. Such antennas are therefore not counted in what follows. $\Xmaxpos$ is at a distance of $71$\,km. The effective opening angle is $\omega_{\rm eff} = 0.52\degree$ (see~\ref{app:degeneracy}). $\delta \Delta t$ is defined as a quarter period of the dominant frequency $f_0 = 80$\,MHz (same frequency used in Eq.~\eqref{eq_heron:fraunhofer}): $\delta \Delta t \approx 3.1$~ns. Larger misalignments would reduce the power of the summed signal by a factor 2. Eq.~\eqref{eq_heron:deltaD} then gives $\delta D \approx 4.7$~km, larger than the typical longitudinal elongation of the shower, around $\sim3$\,km.\footnote{The value of $\sim3$\,km is extracted from the simulations. A quick calculation based on the atmosphere depths gives an emission region upper bound of $\leq5$\,km, using $\Xmax\sim 500$--$600$\,g\,cm$^{-2}$, and the typical air density at these altitudes ($\rho_{\rm air}\sim 10^{-3}$g\,cm$^{-3}$)}

\begin{figure}[!ht]
    \centering
    \includegraphics[width=0.7\textwidth]{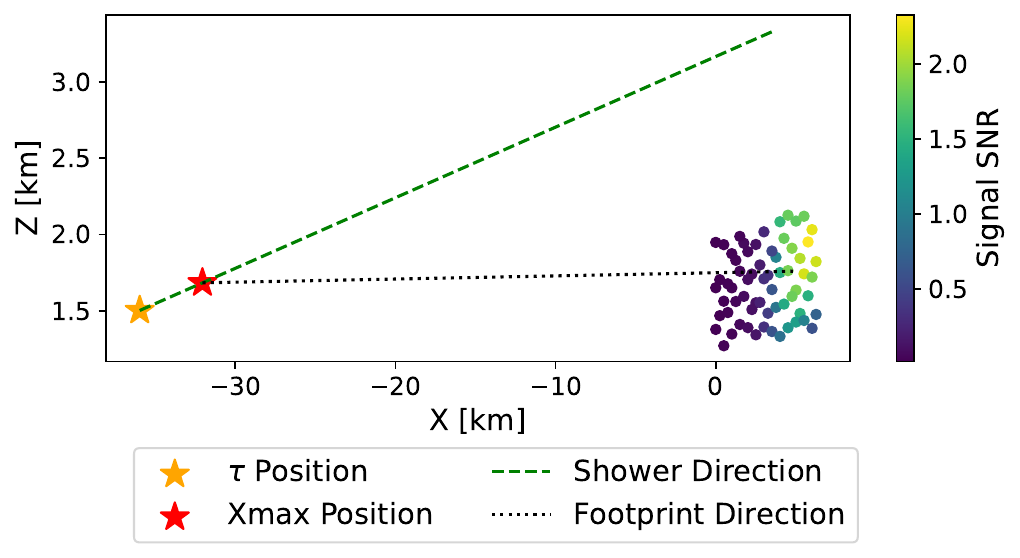}
    \includegraphics[width=0.7\textwidth]{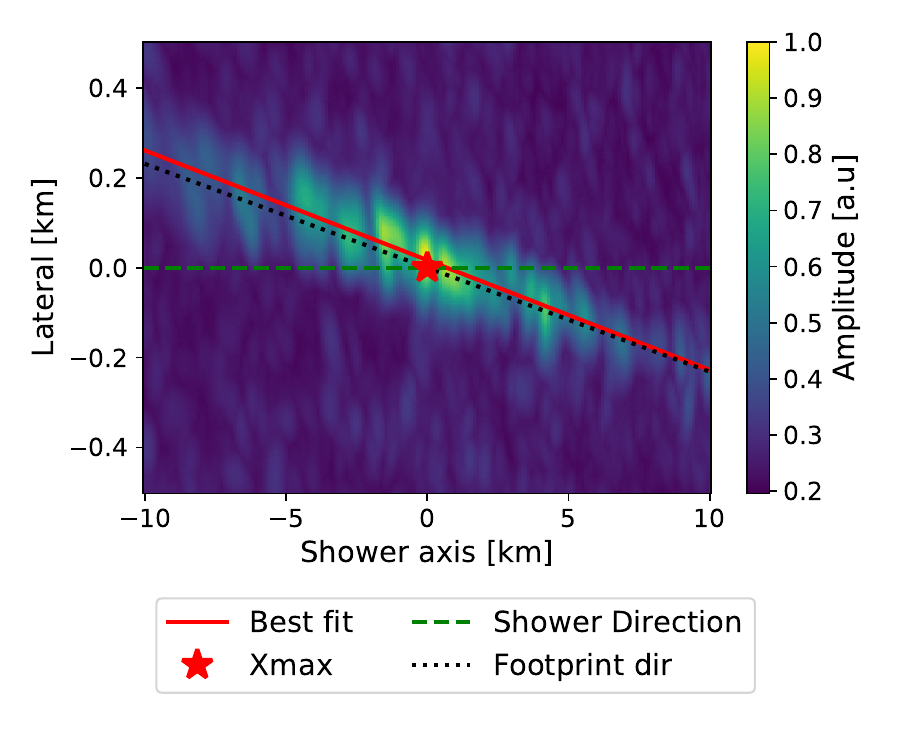}
    \caption{Top: Geometry of the event. The green line represents the shower axis, the black line represents the line of sight direction, i.e. the direction between the average position of the antennas $\rbm{p_b}$ and the shower maximum position $\Xmaxpos$. Bottom: Interferometric map displayed in the vertical plane containing the shower axis. The points with the highest coherence are concentrated along the black line. They align along the line of sight direction rather than the shower axis (green). The reconstructed direction (red) closely follows the line of sight direction.}\label{fig_heron:example_bias}
\end{figure}

Consequently the points of highest coherence in the interferometric map lie along the line of sight direction rather than the shower axis. This is illustrated in Fig.~\ref{fig_heron:example_bias}: the reconstructed direction (red) follows the black line joining the average position of the antennas to the emission region (approximated by $\Xmaxpos$) rather than the shower axis (green).

\subsection{Performance on simulated events}
\begin{figure}[!ht]
    \centering
    \includegraphics[width=0.7\textwidth]{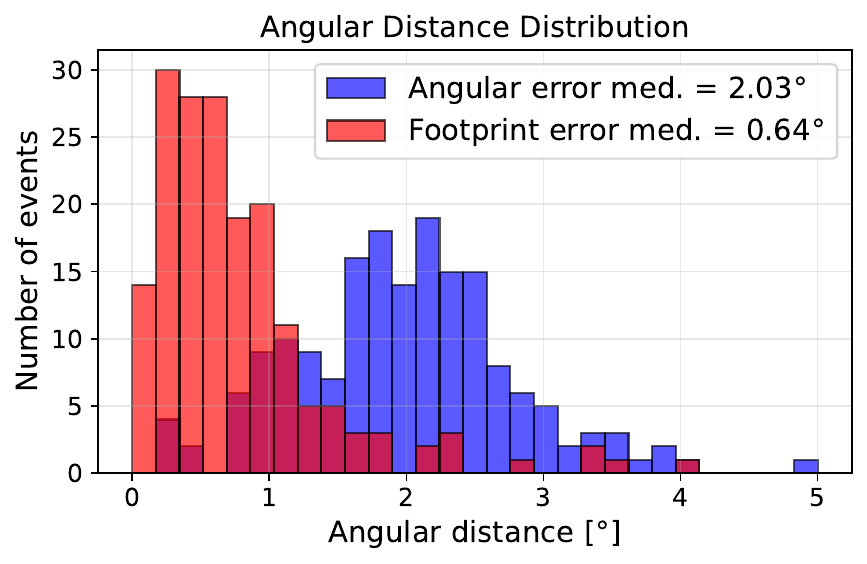}
    \includegraphics[width=0.7\textwidth]{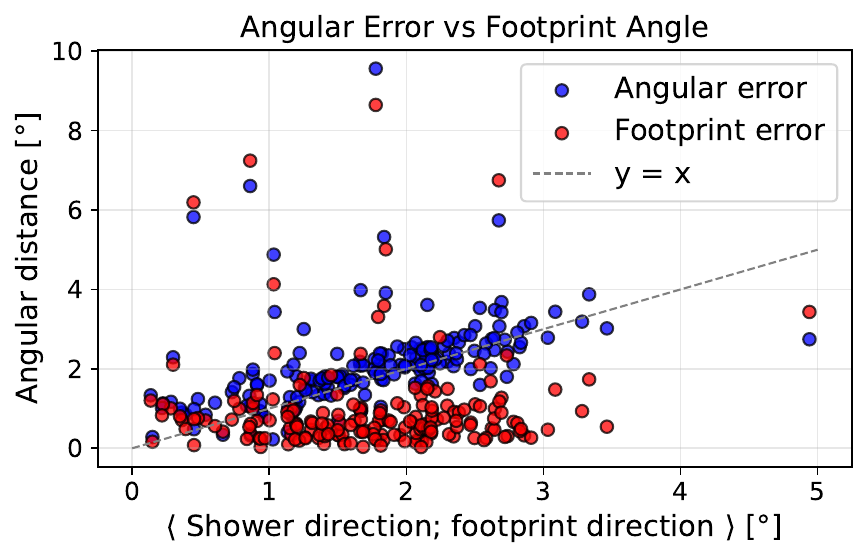}
    \caption{Top: Distribution of angular errors for the interferometric reconstruction. The blue histogram represents the angular error between the reconstructed direction and the true direction, while the red histogram represents the angular distance between the reconstructed direction and the line of sight direction. Bottom: same data, plotted here against the angle between the true direction and the line of sight direction. The angular error is following closely the angle between the true direction and the line of sight direction, confirming the bias of the interferometric reconstruction. }
    \label{fig_heron:interf_bias}
\end{figure}

Applying the interferometric reconstruction to the full simulation set, with gaussian noise and no jitter, makes this bias explicit. Fig.~\ref{fig_heron:interf_bias} (top) shows the distribution of the angular error, defined as the angular distance between the reconstructed and the true direction. Its median, $2.03\degree$, quantifies the angular resolution of the method. Shown in red is the ``line of sight angular error'', i.e. the angular distance between the reconstructed direction and the line of sight direction. It is much smaller, which supports the bias hypothesis. The bottom panel shows that the angular error correlates closely with the angle between the true and the line of sight directions, i.e. with how off-centered the shower is with respect to the array.
A second effect degrades the performance: the method assumes that a source point emits a coherent signal received by all antennas, which is not the case: the emission is beamed along the shower axis, so only antennas within the beam receive a coherent signal. Antennas outside the beam receive a very weak signal, if any; they mostly add noise to the coherent summation and do not improve the global SNR of the summed signal.

The influence of the array aperture and of the limited bandwidth on the interferometric map has been studied in~\cite{OlafApperCorr}. The measured map is modeled as the ideal emission pattern convolved with a kernel encoding the resolution of the technique, computed with MGMR3D~\cite{MGMR3D}; the charge-current parameters (and with them the direction and $\Xmax$) are then recovered by forward-folding this kernel. The method was however demonstrated on low-inclination cosmic rays measured by dense arrays in a noise-free environment, and its extension to sparse arrays and near-horizontal showers remains to be established.

The next section describes a method that reconstructs both the direction and a proxy of $\Xmaxpos$ by adapting the interferometric reconstruction so as to bypass the bias described above.

\section{ADF-interferometry}\label{sec:adfbf}
Three observations follow from the study presented in Sec.~\ref{sec:limits}:
\begin{enumerate}
    \item the reconstructed direction is biased towards the antennas with signal in the array;
    \item many antennas receive no signal and add noise only to the coherent summation;
    \item interferometry is based on timing information only, while classical reconstruction techniques exploit the characteristic features of the radio emission, such as the Cherenkov enhancement or the geomagnetic asymmetry.
\end{enumerate}
Methods have been proposed to mitigate the second issue, for instance in ongoing work on SKA-Low~\cite{Saha:ARENA2026}, where shower-core positions are sampled in addition to source points and only antennas within a given radius of the core are retained. This does not address the first problem, so we exploit instead the third point and weight each antenna by the true SNR expected on it for the event at hand. That SNR is not measurable, since the signal is usually buried in the noise, but it can be predicted from the shower geometry, at the cost of two additional parameters to be inferred, the zenith angle $\theta$ and the azimuth angle $\phi$ of the direction of origin of the shower in the reference frame of the detector. Eq.~\eqref{eq_heron:standard_bf} then becomes
\begin{align}\label{eq_heron:weighted_bf}
    \rbm S \left( \theta,\, \phi,\, \rbm{X_s};\ t \right) = \sum_i w_i(\theta,\, \phi,\, \rbm{X_s}) \rbm{S}_i\left( t + \Delta t_i (\rbm{X_s}) \right),
\end{align}
where the weights $w_i$ depend on the antenna and on the scanned parameters, and are computed from the expected shower geometry.

\subsection{Weighted interferometry} \label{sec:weights}
\begin{figure}[!ht]
    \centering
    \includegraphics[width=0.8\textwidth]{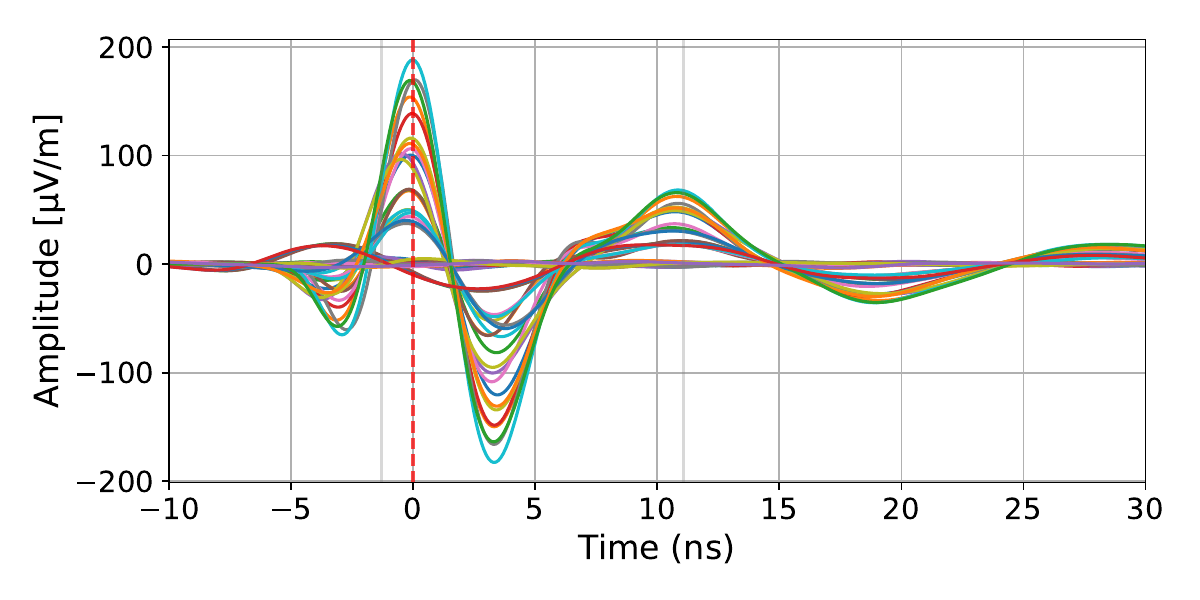}
    \caption{Illustration of the alignment of the signal peaks when the time delays are computed from the position of the shower maximum $\Xmaxpos$.}\label{fig_heron:allignment}
\end{figure}

We write $\rbm{X_e}$ for the effective radio emission point~\cite{Decoene:2022}, which coincides with $\Xmaxpos$ to within the length of the emission region, and $(\theta_0,\, \phi_0)$ for the true direction of the shower. The optimal weights are those for which the amplitude of $\rbm S$ is maximal at $\rbm{X_s} = \rbm{X_e}$, $\theta = \theta_0$ and $\phi = \phi_0$.
We make in the following two strong hypotheses:
\begin{itemize}
    \item signal peaks align and add coherently when the time delays are computed from $\rbm{X_s} = \rbm{X_e} \approx \Xmaxpos$. The validity of this hypothesis is illustrated in Fig.~\ref{fig_heron:allignment}, where the $65$ antenna signals are aligned within an RMS of $2.1$\,ns once corrected for the time delays from $\Xmaxpos$.
    \item The radio signal amplitude can be successfully described by the Angular Distribution Function (ADF) profile introduced in~\cite{GUELFAND2025} and briefly summarized in~\ref{app:adf}. This hypothesis is validated in~\cite{Guelfand:ICRC2025}.
\end{itemize}

Following these two hypotheses, the signal on antenna $i$ can then be written
\begin{align}\label{eq:signal}
    \rbm{S}_i(t) = \mathcal{A}_i\,\rbm{s}_i(t-\Delta t_i(\rbm{X_e})) \ ,
\end{align}
where $\mathcal{A}_i$ is the ADF amplitude predicted at antenna $i$ from $\rbm{X_e}$ for angles $\theta_0$ and $\phi_0$, and $\rbm s_i$ is a dimensionless waveform with a peak amplitude of $1.0$ at $t=0$~ns.
We show in~\ref{app:snr_optimality} that, under these hypotheses, the SNR of the coherent signal is maximized by the normalized ADF amplitudes
\begin{align}\label{eq_heron:weights}
    w_i(\theta,\, \phi,\, \rbm{X_s}) = \frac{f^{\rm ADF}\left( \rbm{p}_i, \rbm{X_s}, \theta, \phi, A=1 \right)}{\left\| f^{\rm ADF}\left( \rbm{P}, \rbm{X_s}, \theta, \phi, A=1 \right)\right\|} \ ,
\end{align}
where $\rbm{P}$ is the vector of all antenna positions. The term in the denominator is the norm of the vector of ADF values over all antennas. This maximum is reached if and only if $\rbm{X_s} = \rbm{X_e}$, $\theta=\theta_0$ and $\phi=\phi_0$, including in the presence of noise, provided the coherent signal is strong enough to be detected (Eq.~\eqref{eq_heron:recons_validity_condition}). The metric to be maximized for a successful reconstruction is then
\begin{align}
    \mathcal{M}(\theta,\, \phi, \rbm{X_s}) = \max_t \qty(\norm{\sum_i w_i(\theta, \, \phi,\, \rbm{X_s})\, \rbm{S}_i\qty( t + \Delta t_i (\rbm{X_s}))}) \ .
\end{align}
Scanning over $\theta$, $\phi$ and $\rbm{X_s}$ parameter space thus provides a joint reconstruction of the direction and of the source position. Note that the dependency of the weights on the ADF scaling factor $A$ cancels out through the normalization in Eq.~\eqref{eq_heron:weights}, hence amplitude calibration is not required at this stage.

\subsection{Tuning of the ADF parameters}\label{sec:adftuning}
The ADF model describes the signal amplitude at each antenna as a function of its angle $\omega$ to the shower axis, its azimuthal angle $\eta$, the geomagnetic angle $\alpha$ and the distance $l$ to the shower maximum. Beyond the overall scaling factor $A$ (irrelevant here because of weight normalization, see Eq.~\eqref{eq_heron:weights}) it depends on the width $\delta \omega$ of the Cherenkov ring, on the geomagnetic asymmetry strength $G_A$, and on the Cherenkov angle $\omega_c$. The three are set as follows.

\paragraph{Cherenkov angle $\boldsymbol{\omega_c}$} In the general ADF formulation, $\omega_c$ is computed with a toy model that evaluates the time delay between two points on the shower axis and the antenna, and returns the angle minimizing this delay. This refinement is unnecessary here, as the shower develops almost parallel to the ground and the radio signals travel over short distances at nearly constant altitude. 
The signal therefore experiences an approximately constant refractive index $n$, computed at the source location, throughout its propagation, so that the classical approximation
\begin{align}
    {\omega_c} = \arccos\left(\frac{1}{n}\right)
\end{align}
holds.
We found a typical Cherenkov angle $\omega_c\approx1.4\degree$ since the refractivity at our altitude is around $n-1\approx 3\times10^{-4}$. This is significantly different from cosmic rays because the refractivity at $\Xmax$ altitude for them is around $n-1\approx 1\times10^{-4}$ leading to a much narrower angle $\sim 0.8\degree$.

\paragraph{Cone width $\boldsymbol{\delta \omega}$} Fitting a large subset of events shows that a value $\delta \omega = 1.2$ describes the simulated signals well, this corresponds to a ring of width $\Delta \omega \approx 0.6\degree$ in the reformulation of~\ref{app:reformulation}.

\paragraph{Geomagnetic asymmetry $\boldsymbol{G_A}$} This parameter must be tuned for each site, as it depends on the altitude of the detector and on the strength of the local geomagnetic field, and it may also depend on $\theta$.
ADF adjustment of the events from our simulation dataset shows that a constant value $G_A=0.5$ can be chosen for the site considered for this study. It is worth noting that for an event forming an angle $\alpha=30\degree$ with the geomagnetic field, Eq.~\eqref{eq:geomag} gives $f^{\rm Geom} = 1 + \cos(\eta)$, so that the geomagnetic emission is then entirely canceled by the charge-excess emission for an antenna at $\eta=180\degree$.

\subsection{Illustration on simulated events}\label{sec:illustration}
\begin{figure}[!ht]
    \centering
    \includegraphics[width=0.95\textwidth]{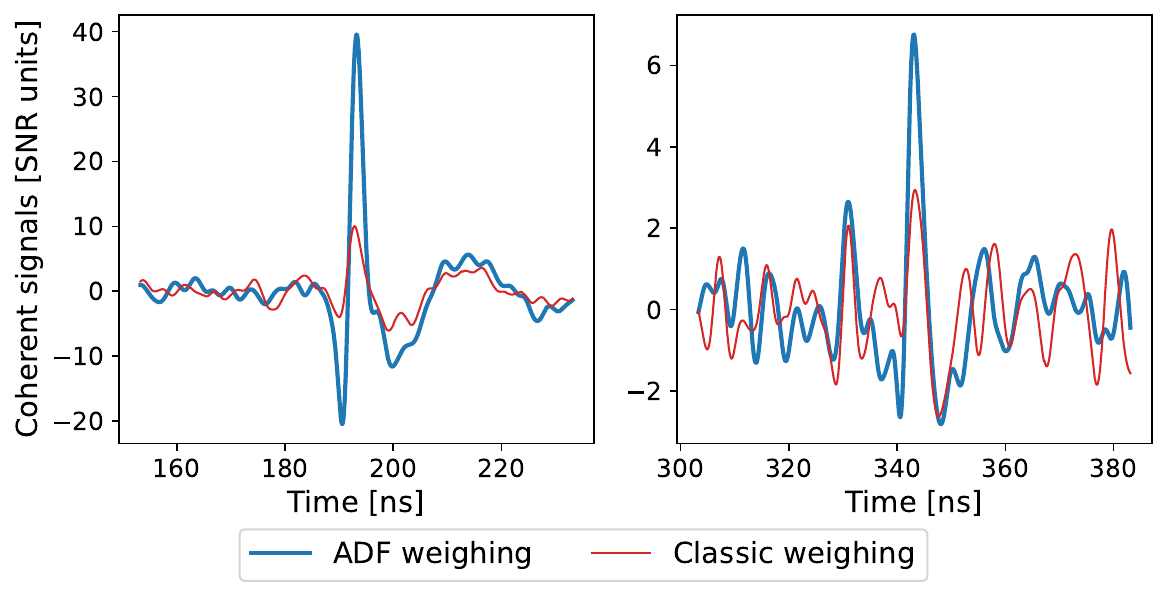}
    \caption{Coherent waveforms with the ADF-weighted sum (blue) compared with the classical interferometric summed signal (red), for two different events from the simulation dataset. Interferometric sum is computed with true direction and $\rbm{X_s} = \Xmaxpos$. The SNR of the ADF-weighted summed signal is higher by a factor larger than 2. For the second event (right), the ADF-weighting brings the peak out of the noise.}\label{fig_heron:SNR_optimization}
\end{figure}

The improvement brought by the ADF weights in the interferometry treatment (i.e. using Eq.~\eqref{eq_heron:weighted_bf} instead of \eqref{eq_heron:standard_bf}) is visible directly on the coherent waveforms. In Fig.~\ref{fig_heron:SNR_optimization} one can see that the SNR improves by a factor larger than $2$ with the ADF-weighted sum for these two events.
\begin{figure}[!ht]
    \centering
    \includegraphics[width=0.75\textwidth]{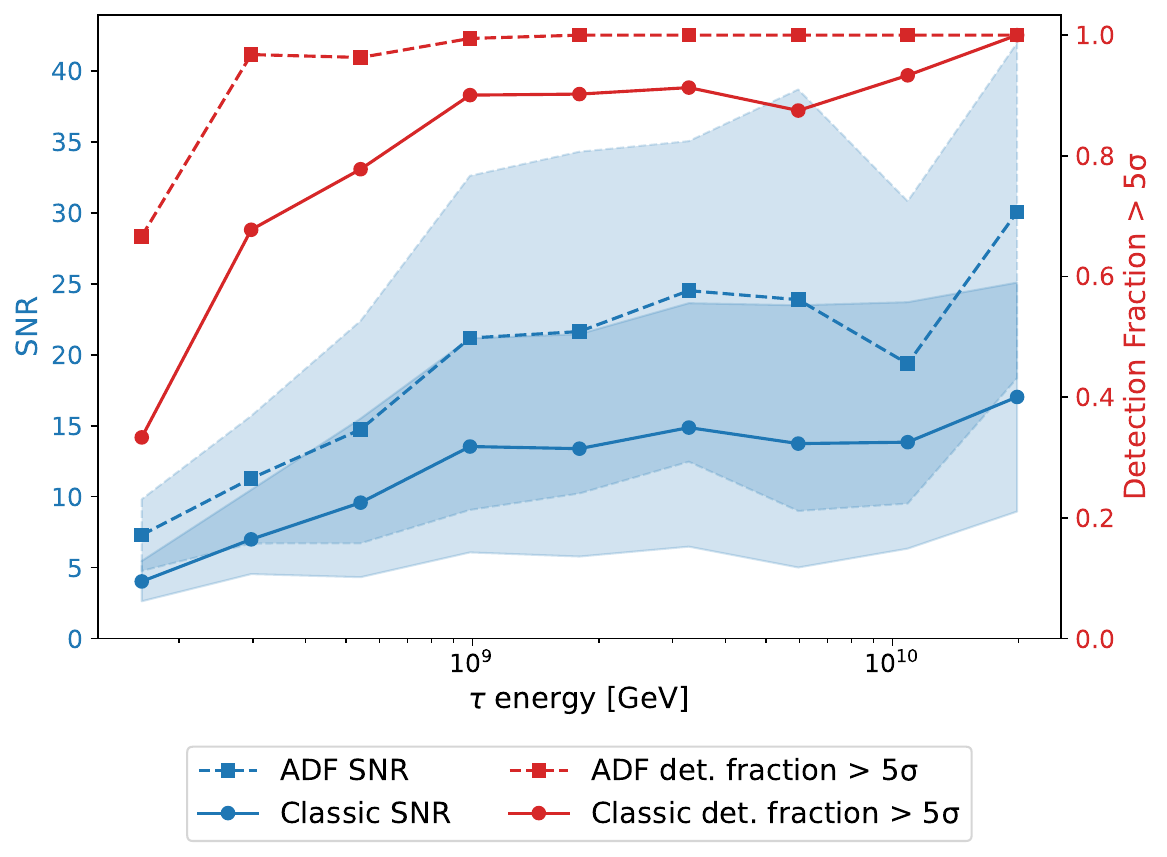}
    \caption{Evolution of the observed SNR with the $\tau$ lepton energy. The ADF-weighted summed signal (dotted) has a higher SNR than the classical interferometric summed signal (solid) at all energies. The detection fraction, defined as the fraction of events with an SNR above $5$, is also systematically higher for the ADF-weighted summed signal. Only triggered events are represented.}
    \label{fig_heron:snr_vs_energy}
\end{figure}
A more systematic study shows that this improvement holds over the whole energy range with a factor of $\sim 1.7$ (see Fig.~\ref{fig_heron:snr_vs_energy}). It translates into a higher detection fraction (the fraction of events whose coherent SNR exceeds $5$, i.e. for which the signal is clearly visible in the summed signal and thus reconstruction can be performed) and therefore lowers the energy threshold above which an event can be reconstructed.

\begin{figure}[!ht]
    \centering
    \includegraphics[width=0.98\textwidth]{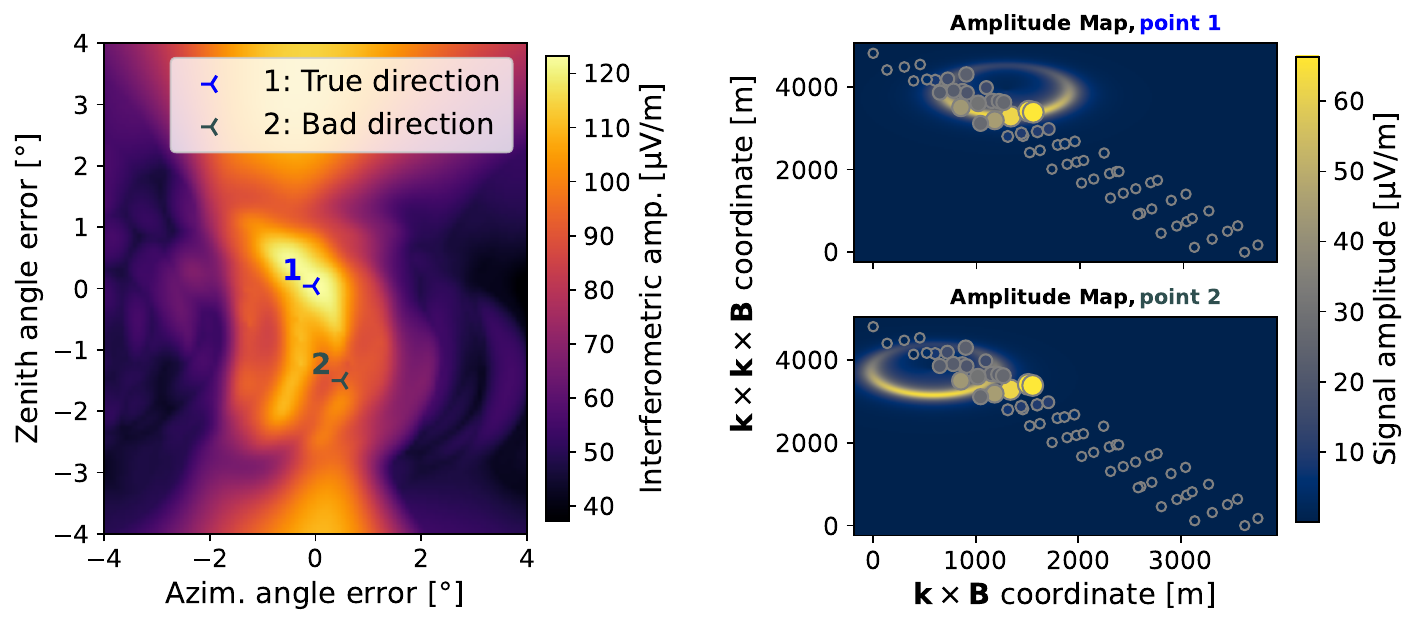}
    \caption{Left: Map of observed amplitude as a function of the scanned direction (x-axis:$\phi$ offset, y-axis:$\theta$ offset). The true direction is at the center of the map. The reconstructed direction is defined as the direction that maximizes the observed amplitude. Top right: Amplitude pattern on individual antennas for the reconstructed direction (blue point 1). Bottom right: Amplitude pattern on individual antennas for a point with lower observed SNR (black point 2). The background shows the expected amplitude from the ADF model. Here again we use $\rbm{X_s} = \Xmaxpos$. }\label{fig_heron:interf_adf_map}
\end{figure}

The behavior of the coherence metric itself is illustrated in Fig.~\ref{fig_heron:interf_adf_map}, for an event of true direction $\theta=92.5\degree$, $\phi=180\degree$ and $\tau$ total energy $E_\tau=10^{17.5}$~eV. Scanning the zenith and azimuth angles on a grid within $4\degree$ of the true direction ($100\times100$ points resulting in a $0.08\degree$ step) produces a map of the observed amplitude of the summed signal. A clear maximum appears at the center of the map, i.e. at zero offset from the true direction. At that point (blue point 1) the true SNR measured on the individual antennas matches the amplitude pattern predicted by the ADF model; at a point of lower observed SNR (black point 2) it does not. The maximum of the map therefore coincides with the configuration in which the predicted and observed footprints agree, which is precisely the behavior established in~\ref{app:snr_optimality}.

\subsection{Regularization terms}\label{sec:priors}
As Fig.~\ref{fig_heron:interf_adf_map} shows, these maps are not convex and exhibit several local maxima, the largest one not always being the closest to the true direction. Two regularization terms are added to mitigate this difficulty.

We first encode the prior knowledge that the shower is upward-going, i.e. $\theta>90\degree$:
\begin{align}
    p_{\rm up}(\theta) = a_{\rm up} \left(\frac{1}{1+\exp(-10(\theta-89.8^\circ))}-1\right).
\end{align}
with $\theta$ in degrees. For $\theta<89.8$° (downward going showers), the reguarization term tends to $-a_{\rm up}$. For $\theta>90$° (upward going showers) the regularization tends to $0$. 
The amplitude is chosen as $a_{\rm up} = 100\,\mu$V\,m$^{-1}$. This is about eight times the noise level of the coherent signal ($\sigma_{\rm noise} = 13\,\mu$V\,m$^{-1}$, see~\ref{app:snr_optimality}), large enough to suppress the spurious maxima corresponding to downward-going showers. It does not bias the reconstruction of genuine upward-going events, because the penalty decays as $e^{-10\Delta\theta}$ above the cutoff: it is already below $2\,\mu$V\,m$^{-1}$ at $0.4\degree$ above it, i.e. below $2\%$ of a typical value of $\mathcal{M}$. We verified that varying $a_{\rm up}$ by a factor of two in either direction leaves the angular resolution unchanged. The cutoff is set to $89.8\degree$ rather than $90\degree$ to avoid biasing showers going slightly down, a possible trajectory in a mountain area.

The second regularization penalizes solutions where the shower axis lies too far from the array, an unlikely scenario. For every antenna $i$ we compute $\Delta \omega_{c,\,i}$, the angular distance between $\omega_i$ and the local Cherenkov angle $\omega_{c,\,i}$, and define
\begin{align}
    p_{\rm c}(\theta,\, \phi,\, \rbm{X_s}) = - a_{\rm c}\sum_i w_i(\theta, \, \phi,\, \rbm{X_s}) \Delta \omega_{c,\,i}^2,
\end{align}
with $a_{\rm c}=1000~\mu$V\,m$^{-1}$\,rad$^{-2}$. Since the weights are normalized, this term favors configurations in which the largest weights fall on antennas with a small $\Delta \omega_{c,\,i}$, i.e. it encodes the expectation that the most intense antennas are likely to be close to the Cherenkov ring.

Strictly speaking these are not priors but ad-hoc regularization terms, whose role is to keep the reconstruction away from nonphysical solutions, but for simplicity we will refer to them as priors in the following. The final quantity to maximize is
\begin{align}\label{eq_heron:target_function_reg}
    \mathcal{M}_{\rm reg}(\theta,\, \phi,\, \rbm{X_s}) = \mathcal{M}(\theta,\, \phi,\, \rbm{X_s}) + p_{\rm up}(\theta) + p_{\rm c}(\theta,\, \phi,\, \rbm{X_s}).
\end{align}
%

\section{Performance}\label{sec:perf}
We now assess the performance of the ADF-interferometry, first using the true value of $\Xmaxpos$ (Sec.~\ref{sec:proof_principle}), then in the realistic case where the direction and $\rbm{X_e}$ (see Sec.~\ref{sec:weights}) are reconstructed jointly and compared to $\Xmaxpos$ (Sec.~\ref{sec:joint}). The energy reconstruction is addressed in Sec.~\ref{sec:energy}.

\subsection{Proof of principle: reconstruction with known \texorpdfstring{$\Xmaxpos$}{Xmax}}\label{sec:proof_principle}
\begin{figure}[!ht]
    \centering
    \includegraphics[width=0.65\textwidth]{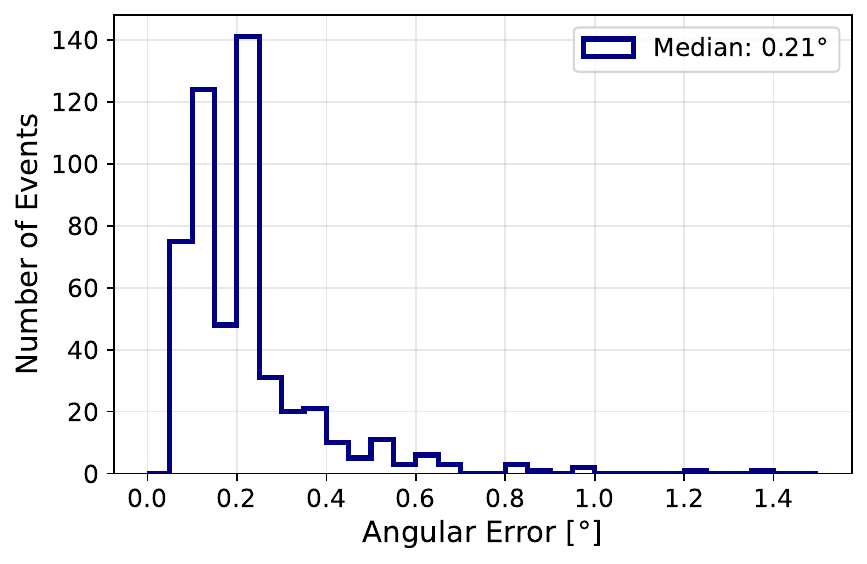}
    \caption{Histogram of angular errors for the 548 triggered events without noise or time jitter.}
    \label{fig_heron:angular_error_fromxmax_hist}
\end{figure}

The reconstructed direction is defined as the direction maximizing $\mathcal{M}_{\rm reg}$ as defined in Eq.~\eqref{eq_heron:target_function_reg} over the grid described in Sec.~\ref{sec:illustration}. Applied to the noise-free waveforms and using the regularization terms described earlier, this yields a median angular error of $0.21\degree$ (see Fig.~\ref{fig_heron:angular_error_fromxmax_hist}). This result is not grid-limited since the step is $0.08\degree$. The ADF-interferometry therefore succeeds in correcting for the bias identified in Sec.~\ref{sec:degeneracy}, improving the resolution from $2.03\degree$ to $0.21\degree$, assuming known position of $\Xmaxpos$.
\subsubsection{Effect of noise}
\begin{figure}[!ht]
    \centering
    \includegraphics[width=0.65\textwidth]{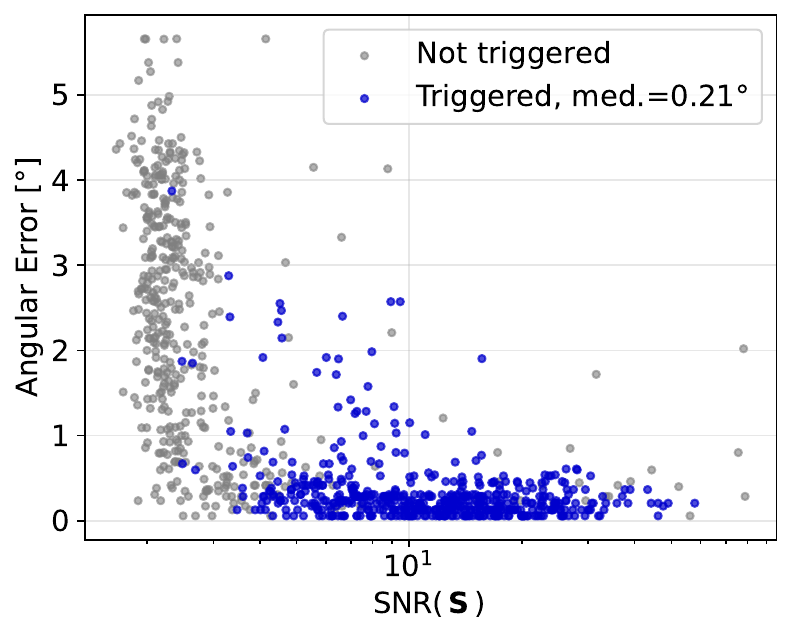}
    \caption{Angular errors for the $972$ events against the measured SNR of the summed signal on noisy simulations. The color indicates whether the event triggered ($548$ in total) on one of the phased stations (blue) or not (gray). }
    \label{fig_heron:scatter_angular_snr}
\end{figure}

Noise mimicking the Galactic contribution (see Sec.~\ref{sec:simu}) is now added to the simulated waveforms. Fig.~\ref{fig_heron:scatter_angular_snr} shows the angular errors of the 972 simulated events against the observed SNR of the coherent signal. The angular errors remain below $1\degree$ for most triggered events (i.e., $\mathrm{SNR}_{\rm phased} \geq 4$ corresponding to an amplitude of $10.6\,\mu$V\,m$^{-1}$ on a single antenna, see Sec.~\ref{sec:simu}) and for all of those with $\mathrm{SNR}(\,\rbm{S}\,)>10$, which is consistent with the condition $\mathcal{M} \gg \sigma$ defined in Eq.~\eqref{eq_heron:recons_validity_condition}. The median angular error for triggered events is $0.21\degree$, i.e. unchanged with respect to the noise-free case.

\begin{figure}[!ht]
    \centering
    \includegraphics[width=0.98\textwidth]{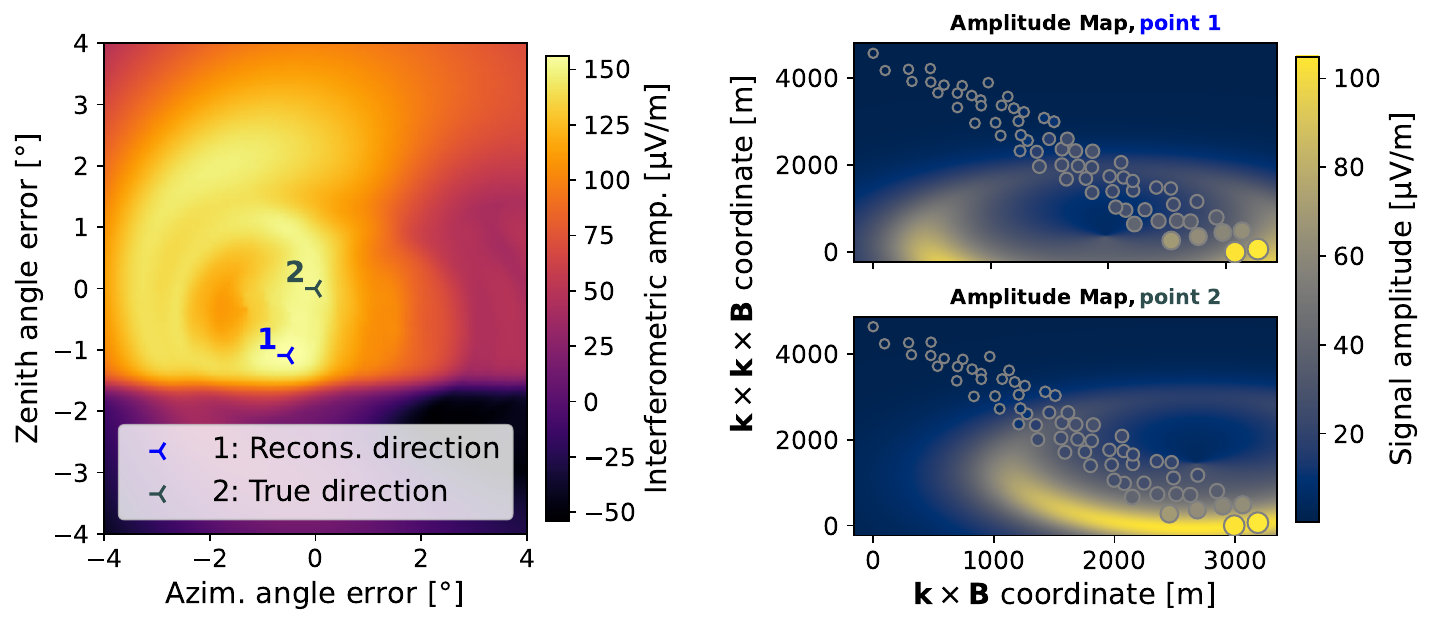}
    \caption{Example of an outlier event. Left: Map of observed amplitude as a function of the scanned direction, with the true direction (point 2) at the center of the map and the reconstructed direction (point 1) at the maximum of observed SNR. Right: Amplitude pattern on individual antennas for the two points. The background shows the expected amplitude from the ADF model. The observed footprint is compatible with two different geometries, leading to a degeneracy in the reconstruction.}\label{fig_heron:interf_adf_map_noise_outlier}
\end{figure}

Outliers appear only below $\mathrm{SNR}(\,\rbm{S}\,)=10$. Visual inspection shows that only a small fraction of the footprint is detected for these events, so that several distinct footprints remain compatible with the observation. Fig.~\ref{fig_heron:interf_adf_map_noise_outlier} illustrates one such case: two well-separated maxima are present in the map, and the reconstruction settles on the wrong one. This degeneracy is directly visible on the map, and may therefore be used as a quality criterion to reject unreliable events in a future treatment.

\subsubsection{Effect of time resolution}\label{sec:jitter}
\begin{figure}[!ht]
    \centering
    \includegraphics[width=1\textwidth]{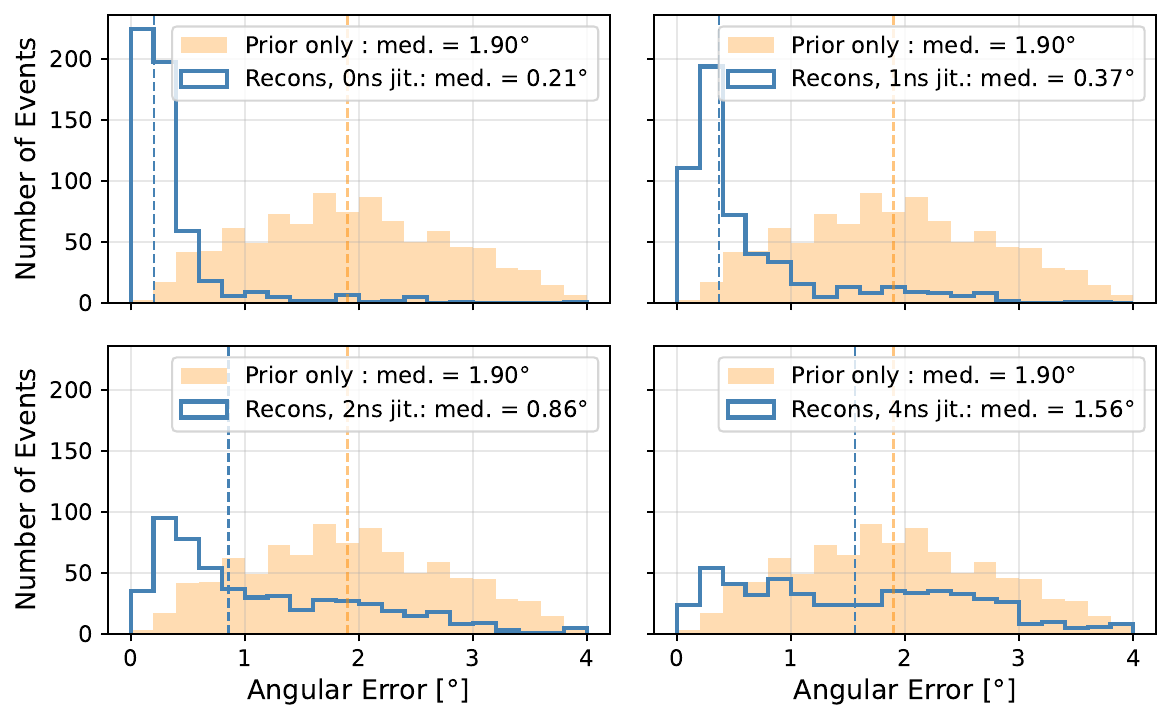}
    \includegraphics[width=.75\textwidth]{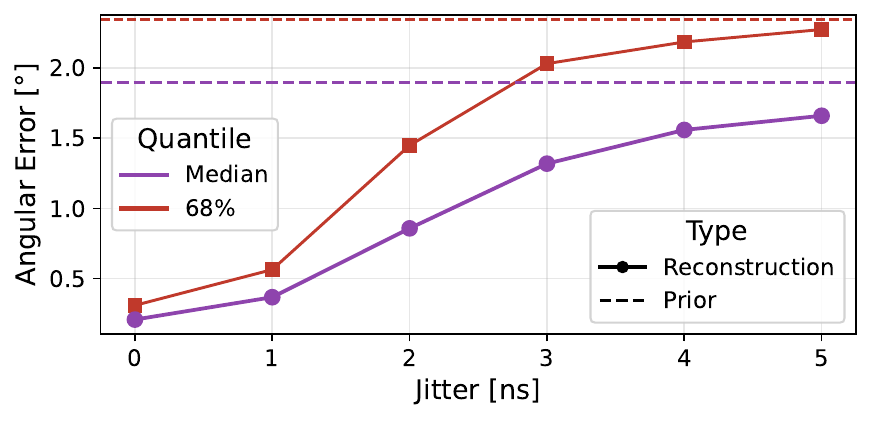}
    \caption{Top: Histograms of angular errors for the 548 triggered events for different time jitters. The median angular error is 0.21$\degree$ for $0$\,ns, and goes to $1.18\degree$ for $5$\,ns. The histograms of angular errors are also shown for the reconstruction using the priors only. Bottom: evolution of the median angular error and $68\%$ quantile of the angular error distribution as a function of the time jitter. The values for the priors only reconstruction are also shown as horizontal lines.}
    \label{fig_heron:jitter_resolution_histograms}
\end{figure}

We then add to the antenna times a random time jitter drawn from a zero-mean Gaussian distribition with a standard deviation $\sigma_{\rm t}$ ranging from 0 to $5$\,ns to mimic timing uncertainty. As expected, the reconstruction performance degrades as $\sigma_{\rm t}$ increases (Fig.~\ref{fig_heron:jitter_resolution_histograms}). At large $\sigma_{\rm t}$ values the distribution of angular errors approaches the one obtained with the regularization terms alone (orange), indicating that the coherent summation no longer contributes and that the result is driven by the priors. This is the expected behavior: once the jitter exceeds the typical width of the signal (around $5$\,ns) the transient signals no longer align and coherent summation simply fails. Below that threshold, the critical value of time jitter will depend on whether one aims for a sub-degree angular resolution or a sub-kilometer shower location reconstruction.

The angular resolution is already degraded at $\sigma_{\rm t} = 1$\,ns, whereas a sub-degree angular resolution is still reached at $\sigma_{\rm t} = 3$\,ns. These results are consistent with several prior studies~\cite{FrankInterfArray,beaconAuger,interfvjitter} setting stringent constraints (yet within reach, see e.g.~\cite{PierreAuger:timing}) on time synchronization between antennas to achieve a sub-degree angular resolution.

\subsection{Joint reconstruction of the direction and of \texorpdfstring{$\Xmaxpos$}{Xmax}}\label{sec:joint}
In the previous sections we assessed the potential for the ADF-interferometry, using the true $\Xmaxpos$ value (in place of $\rbm{X_e}$ formally) in the interferometric treatment. In practice the direction and $\Xmaxpos$ must be reconstructed jointly. We do so by global minimization of $-\mathcal{M}_{\rm reg}$, defined in Eq.~\eqref{eq_heron:target_function_reg}, over the five-dimensional parameter space $(\theta,\, \phi,\, \rbm{X_s})$. Reconstructing the full map instead, on a grid or by Monte Carlo sampling, would give a richer picture of the parameter space at a much higher computational cost; we return to this in Sec.~\ref{sec:conclusion}.

A gradient-descent method is ill-suited, as the function is not convex and exhibits many local minima (see e.g. Fig.~\ref{fig_heron:interf_adf_map_noise_outlier}). We therefore use a differential evolution algorithm~\cite{diff_evol}, which iteratively refines a set of candidate solutions over successive generations and evaluates their fitness with $\mathcal{M}_{\rm reg}$ until convergence; the same algorithm is used for spherical-wavefront reconstruction in~\cite{GUELFAND2025}. It requires bounds on the parameters: we use $4\degree$ around the true direction for the zenith and azimuth angles, and $20$\,km around $\Xmaxpos$ for the coordinates of $\rbm{X_s}$, with the additional requirement that $\rbm{X_s}$ lie above the ground $h_{\rm g}$: $z> h_{\rm g}$. Here the bounds emulate a phased-station seed whose accuracy is assumed to be better than $4\degree$ in direction and $20$\,km in $\Xmaxpos$ position~\cite{BEACON:2025qcq}. The sensitivity to this assumption is left to a subsequent study.
A population of 90 candidates is generally sufficient; occasionally the global minimum is missed because the population did not explore the full parameter space.

\begin{figure}[!ht]
    \centering
    \includegraphics[width=1\textwidth]{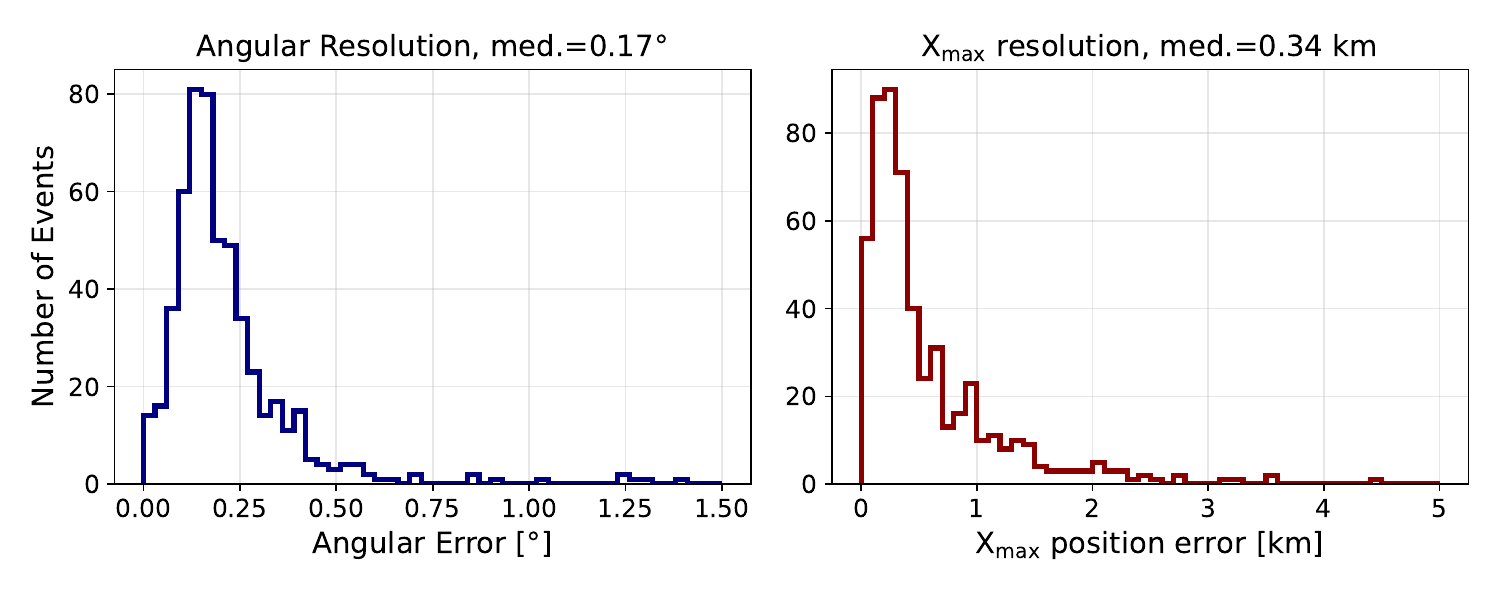}
    \caption{Left: Histogram of angular errors for the 548 triggered events. The median angular error is 0.17$\degree$. Right: Histogram of errors on the reconstructed position $\Xmaxpos$. The median error on the reconstructed position is 0.34~km. This reconstruction is performed with no noise and no jitter.}\label{fig_heron:global_recons_no_noise}
\end{figure}

\begin{figure}[!ht]
    \centering
    \includegraphics[width=0.95\textwidth]{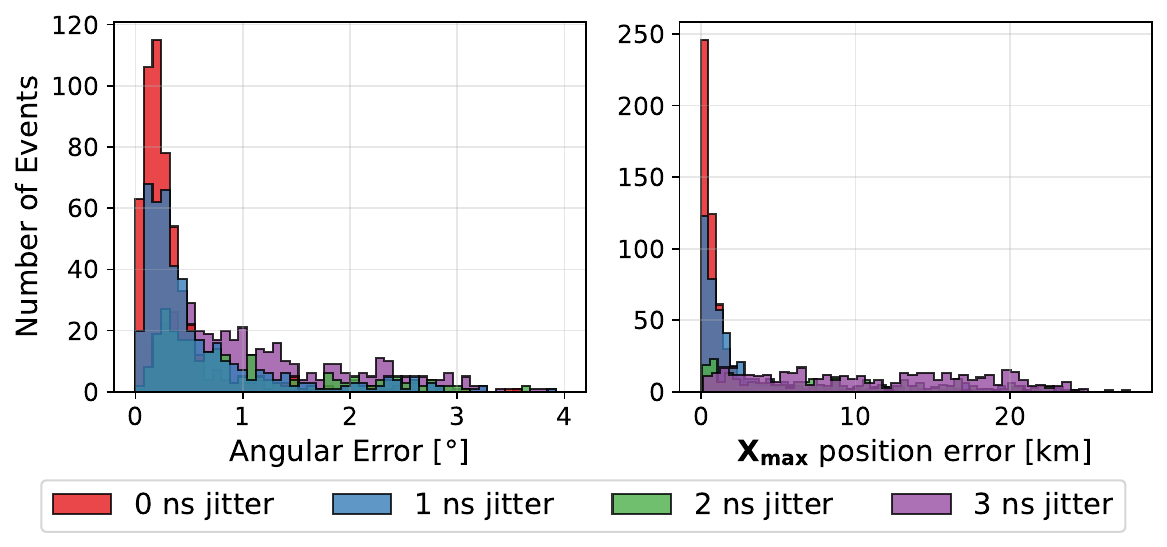}
    \caption{Left: Histogram of angular errors for the 548 triggered events with Gaussian noise and time jitters from 0~ns (red) to 3~ns (purple). Right: Histogram of errors on the reconstructed position $\Xmaxpos$ for the 548 triggered events with Gaussian noise and time jitters from 0~ns (red) to 3~ns (purple). Median values are summarized in Table~\ref{tab:performance}.}
    \label{fig_heron:global_recons_noise_jitter}
    \vspace{1em}
    \begin{tabular}{lcc}
    \hline
    Time jitter & Median angular error & Median $\Xmaxpos$ error \\
    \hline
        0~ns (no noise) & $0.17\degree$ & 0.34~km \\
        0~ns (red)           & $0.23\degree$ & 0.54~km  \\ 
        1~ns (blue)           & $0.34\degree$ & 1.15~km  \\ 
        2~ns (green)           & $0.71\degree$ & 7.43~km  \\ 
        3~ns (purple)           & $0.96\degree$ & 10.71~km  \\ 
    \hline
    \end{tabular}
    \captionof{table}{Performance of the joint reconstruction of the direction and of $\Xmaxpos$ for the 548 triggered events, as a function of the time jitter. All rows include Gaussian noise except the first.}
    \label{tab:performance}
\end{figure}

Fig.~\ref{fig_heron:global_recons_no_noise} displays the results where the median angular error is $0.17\degree$ using the clean waveforms (no noise, $\sigma_{\rm t}$~=~0ns), an improvement over the $0.21\degree$ obtained with true $\Xmaxpos$. This can be explained by two effects: allowing a slight offset on the position of $\Xmaxpos$ can improve the coherence since $\rbm{X_e}$ does not exactly correspond to $\Xmaxpos$, and the previous reconstruction was restricted to a discrete grid of directions with a $0.08\degree$ angular step. The median error on the reconstructed position is $0.34$\,km (keeping in mind that this value actually includes the intrinsic offset between $\rbm{X_e}$ and $\Xmaxpos$), with very few outliers (only 4 above $4$\,km). The remaining angular outliers are the same events as in Fig.~\ref{fig_heron:angular_error_fromxmax_hist}, plus a few cases where the global minimum was simply not found. These are convergence failures rather than a limitation of the method: $\mathcal{M}_{\rm reg}$ evaluated at the true parameters exceeds its value at the reconstructed ones. With a larger population, or more computing budget than the laptop used here, they could be recovered.
We now repeat the joint reconstruction with Gaussian noise and a time jitter from $0$ to $3$\,ns, since at $5$\,ns interferometric reconstruction isn't possible (see Sec.~\ref{sec:jitter}). The results are shown in Fig.~\ref{fig_heron:global_recons_noise_jitter} and summarized in Table~\ref{tab:performance}.

With noise but no jitter, the median angular error is $0.23\degree$ and the median error on $\Xmaxpos$ is $0.54$\,km, a slight degradation with respect to the noise-free case. Adding a $\sigma_{\rm t} =1$\,ns jitter brings these median errors to $0.34\degree$ and $1.15$\,km respectively. Beyond $1$\,ns the degradation is fast, reaching $0.71\degree$ and $7.43$\,km at $2$\,ns; the $\Xmaxpos$ resolution deteriorates markedly faster than the angular resolution, as it relies more directly on the curvature of the wavefront and hence on the arrival-time differences. This reinforces the conclusion of Sec.~\ref{sec:jitter}: the inter-antenna time synchronization is a critical aspect of the detector design for interferometric reconstruction.

\subsection{Proof of concept: energy reconstruction}\label{sec:energy}
The reconstruction described so far makes use of the location of the maximum of $\mathcal{M}_{\rm reg}$ only. The actual value at this maximum yet carries information on the energy of the shower. We outline here a proof of concept exploiting this correlation.

The $\tau$ decays through several channels, and the fraction of its energy carried away by the daughter neutrinos varies strongly from decay to decay. We therefore relate the radio signal to the shower energy $E_{\rm shower}$, defined as the total energy of the decay products that initiate a shower.

\begin{figure}[!ht]
    \centering
    \includegraphics[width=1\textwidth]{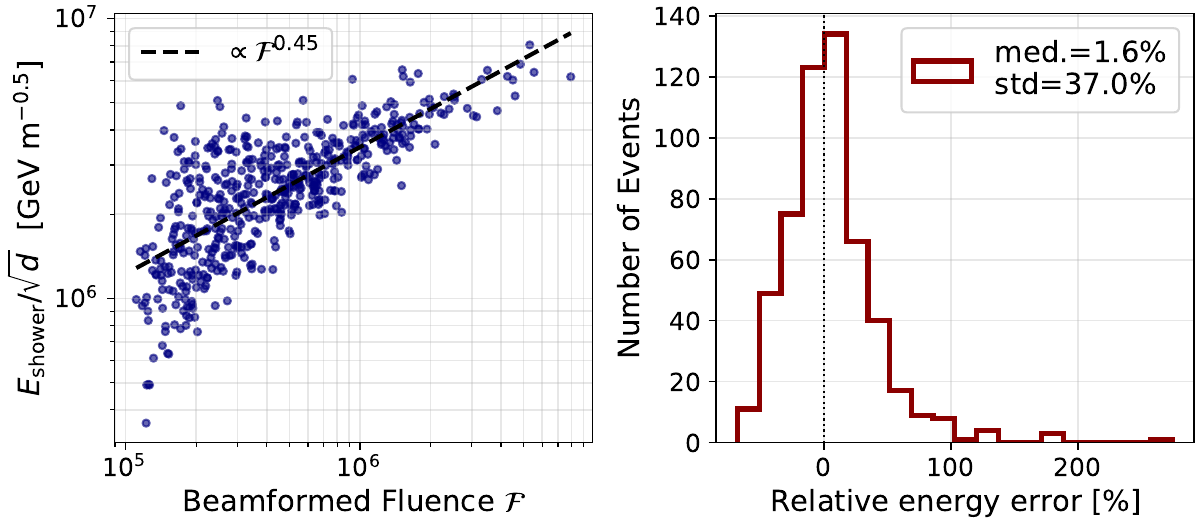}
    \caption{Reconstructed shower energy $E_{\rm shower}$ as a function of the fluence of the summed signal $\mathcal{F}$ and the distance $d$ to the reconstructed $\Xmaxpos$. Left: relation between $E_{\rm shower}/\sqrt{d}$ and $\mathcal{F}$. Right: relative error on the reconstructed energy. The standard deviation of the relative error on the reconstructed energy is $37$\%. This reconstruction has been performed on the noisy signals with no jitter.}\label{fig_heron:energy_reconstruction}
\end{figure}

We fit a simple power law relating $E_{\rm shower}$ to two quantities: the fluence $\mathcal{F}$ of the coherent signal, i.e. the integral of its square over time, which gave slightly better results than the amplitude; and the distance $d$ between $\Xmaxpos$ and the amplitude-weighted barycenter of the antennas. This distance is itself accessible from the reconstruction, being obtained from the estimated $\Xmaxpos$, the reconstructed direction and the amplitudes predicted by the ADF. The fit yields
\begin{align}
    E_{\rm shower} \propto d^{0.5}\, \mathcal{F}^{0.45},
\end{align}
shown in Fig.~\ref{fig_heron:energy_reconstruction} for the 548 triggered events.
The fit is performed in two steps: the distance exponent is adjusted first and then held fixed at $0.5$, and the exponent of the fluence is left free in a second step. The standard deviation of the relative error on the reconstructed energy is $37\%$, showing that the shower energy is recovered with reasonable accuracy. This is only a proof of concept: it could be improved with a more complete model, by targeting the electromagnetic energy of the shower rather than the shower energy, or by exploiting more of the information contained in $\mathcal{M}_{\rm reg}$.
\section{Conclusion and perspectives} \label{sec:conclusion}
We have investigated in this article interferometric reconstruction for low SNR signals induced by inclined EAS in extended radio arrays. We have shown that classical interferometric reconstruction~\cite{HarmInterf} is biased for off-centered events detected by an array subtending a small angle at the source, as the direction is reconstructed along the line of sight direction rather than the shower axis. This results in an angular resolution of median error around $2\degree$. The bias is generic to spherical-wavefront interferometry, while the value quoted here is specific to the selection of the simulated events used to generate the sample (see Sec.~\ref{sec:simu}). 

To overcome this problem, we proposed instead to weight the signal of each antenna by the amplitude predicted by the Angular Distribution Function, which gives an analytical profile of the radio emission for a given source position and direction of propagation. This treatment maximizes the SNR of the coherent signal, and the maximum is reached at the true source position and direction. A joint reconstruction of the direction and $\Xmaxpos$ yields a median error of $0.23\degree$ and of 0.54~km for these two parameters for events including Galactic noise. This is accurate enough for source identification and for neutrino/cosmic-ray discrimination based on $\Xmaxpos$ since $\tau$ induced showers peak at $50$--$100$\,km while very inclined  cosmic-ray induced showers peak at distances larger than $250$\,km. We also show that a relative antenna time synchronization better than $3$\,ns is needed to reach a sub-degree angular resolution, and better than $\sim 1$\,ns to reach the kilometer level for the positioning of $\Xmaxpos$. Finally, a proof of concept of the energy reconstruction yields a $37\%$ standard deviation on the relative error of the reconstructed energy. This treatment could be used in particular in the reconstruction of events detected with the standalone array of the HERON detector.

Several paths could be investigated to improve this method:
\begin{itemize}
    \item The regularization terms can be refined. Rather than a simple prior excluding downward-going events, the Earth crust depth could be evaluated at each iteration to build an actual $\nu_\tau$ interaction probability.
    \item The reconstruction could be applied to voltage signals, with realistic Galactic and instrumental noise and the addition of Radio Frequency Interference.
    \item Rather than ADF, a model of the radiation intensity derived from first principles could be used to estimate the weight associated to each antenna in the interferometric treatment. This would additionally account for the non-spherical curvature of the wavefront, the beaming of the signal and the Cherenkov compression.
    \item The energy reconstruction has not been explored in depth. It could be improved with a more complete model exploiting additional dependencies, as done for the energy reconstruction of cosmic rays with the ADF method~\cite{GUELFAND2025, Guelfand:ICRC2025}.
    \item Only the maximum of $\mathcal{M}$ is used here, whereas classical interferometric reconstruction exploits the full map. Scanning the full five-dimensional space of $\theta$, $\phi$ and $\Xmaxpos$ on a grid or by Monte Carlo sampling could provide a second, independent estimate of the shower energy and direction.
    \item Uncertainties could be estimated by exploring $\mathcal{M}$ with Monte Carlo sampling. Fig.~\ref{fig_heron:interf_adf_map_noise_outlier} suggests that despite a large error on the reconstructed direction, the true direction remains compatible with the degeneracy visible in the map. This would require a proper study of the likelihood function, in particular of the distribution of its maximum over noise realizations.
    \item The study of the full five-dimensional map could also be combined with an adaptation of the finite-aperture correction of~\cite{OlafApperCorr}, to improve the reconstruction of the direction and of the particle profile.
    \item Once a first reconstruction has been performed, a template of the average waveform on the antennas becomes available. The standalone-antenna waveforms could then be cross-correlated with this template (match filtering) and the interferometric reconstruction performed on the resulting signals, allowing a better treatment of the timing uncertainty and a higher SNR on individual antennas. This could be done iteratively, or directly from templates built on a very restricted subset of simulations, typically five~\cite{templatematching}. It would naturally lead to the use of a full signal emulator in the reconstruction.
    \item Finally, the reconstructed direction, energy and possibly particle profile, together with a proper uncertainty estimation, would in principle, open the way to a Bayesian estimate of the decay channel of the $\tau$ lepton, a first step towards a reconstruction of the $\tau$-lepton and, ultimately, of the $\nu_\tau$ energy with rigorous uncertainties.
\end{itemize}
These improvements will be explored in follow-up studies.

\appendix

\section{Geometrical origin of the longitudinal degeneracy}\label{app:degeneracy}
We derive here Eq.~\eqref{eq_heron:deltaD}, which gives the uncertainty on the reconstructed source position along the longitudinal axis, the axis joining the source $\rbm{X_s}$ to a point $\rbm{p_b}$ in the antenna array, close to the average position of the antennas. It will be refined later.

Let $D = \|\rbm{X_s}-\rbm{p_b}\|$ be the distance between the source and $\rbm{p_b}$, and let $\Delta\omega_i$ be the opening angle between $\rbm{p_b}$ and antenna $i$ as seen from $\rbm{X_s}$. For an antenna at a transverse distance $D\,\Delta\omega_i$ from the longitudinal axis, and in the small-angle limit $\Delta\omega_i \ll 1$ appropriate to a sparse array observing a distant source, the propagation distance reads
\begin{align}
    \|\rbm{X_s}-\rbm{p}_i\| \simeq D\left(1 + \frac{\Delta\omega_i^2}{2}\right),
\end{align}
so that, using Eq.~\eqref{eq_heron:delay} with $n \simeq 1$, the arrival-time difference between antenna $i$ and $\rbm{p_b}$ is
\begin{align}\label{eq:app_curvature}
    \Delta t_i \simeq \frac{D\,\Delta\omega_i^2}{2c}.
\end{align}
This is the curvature term that carries all the information on the distance to the source. Differentiating Eq.~\eqref{eq:app_curvature} with respect to $D$ gives the sensitivity of the observable to the source distance,
\begin{align}
    \left|\frac{\partial \Delta t_i}{\partial D}\right| \simeq \frac{\Delta\omega_i^2}{2c}.
\end{align}
An antenna at a small opening angle is therefore almost insensitive to $D$, and the constraint on the source position comes essentially from the antennas at the edge of the array.

Assuming that each arrival-time difference is measured with the same uncertainty $\delta \Delta t$, and that these uncertainties are independent, the antennas combine in quadrature and the variance on $D$ is
\begin{align}
    \frac{1}{\delta D^2} = \sum_i \left(\frac{1}{\delta \Delta t}\frac{\partial \Delta t_i}{\partial D}\right)^2 \simeq \frac{1}{(2c\,\delta \Delta t)^2}\sum_i \Delta\omega_i^4,
\end{align}
which yields Eq.~\eqref{eq_heron:deltaD},
\begin{align}
    \delta D \simeq \frac{2c\,\delta \Delta t}{\sqrt{\sum_i \Delta\omega_i^4}}.
\end{align}
It is convenient to introduce the effective opening angle
\begin{align}
    \omega_{\rm eff} = \left(\frac{1}{N}\sum_i \Delta\omega_i^4\right)^{1/4},
\end{align}
used in Fig.~\ref{fig_heron:uncertainty_computation}, so that $\delta D \approx 2c\,\delta \Delta t/(\sqrt N\,\omega_{\rm eff}^2)$. 
This defines the axis from the source along which the uncertainty is maximal, i.e. $\omega_{\rm eff}$ is minimal. $\rbm{p_b}$ is thus the intersection of this axis with the antenna array. 
The fourth power in $\omega_{\rm eff}$ makes this figure of merit strongly dominated by the outermost antennas, and the quadratic dependence on $\omega_{\rm eff}$ makes the longitudinal uncertainty degrade very quickly as the array subtends a smaller angle. 
\section{The Angular Distribution Function}\label{app:adf}
The ADF is described in detail in~\cite{GUELFAND2025}; we recall here the aspects needed in this paper.

The signal amplitude measured at an antenna location is described by its angle $\omega$ with respect to the shower axis, its azimuthal angle $\eta$ defined in the shower plane $\left(\mathbf{k}\times\mathbf{B},\ \mathbf{k}\times\left(\mathbf{k}\times\mathbf{B}\right),\ \mathbf{k}\right)$, its geomagnetic angle $\alpha$ and its distance $l$ to the shower maximum. In this coordinate system, the signal is given by the phenomenological function
\begin{align}\label{eq:adf}
    f^{\rm ADF}(\omega, \eta, \alpha, l; A, \delta \omega) = \frac{A}{l}\, f^{\rm Geom} (\alpha, \eta, G_A)\, f^{\rm Cherenkov} (\omega; \delta \omega),
\end{align}
where $A$ is a free parameter adjusting the overall amplitude and $f^{\rm Geom}$ describes the geomagnetic asymmetry,
\begin{align}
\label{eq:geomag}
    f^{\rm Geom}(\alpha, \eta, G_{A}) = 1+ G_{A} \frac{\cos(\eta)}{\sin(\alpha)},
\end{align}
with $G_{A}$ the geomagnetic asymmetry strength, parameterized in the general formulation as a function of the zenith angle. The term $f^{\rm Cherenkov}$ is a Lorentzian parametrization of the signal enhancement observed around the Cherenkov angle:
\begin{align} \label{eq:cher}
    f^{\rm Cherenkov}(\omega; \delta \omega) = \frac{1}{1+4 \left[ \dfrac{(\tan(\omega)/ \tan(\omega_{c}))^{2} -1}{\delta \omega}  \right]^{2}},
\end{align}
where $\omega_{c}$ is the Cherenkov angle, computed either numerically from the model of~\cite{GUELFAND2025} or analytically following~\cite{Decoene:2025fmv}, and $\delta \omega$ is a free parameter describing the width of the Cherenkov ring.

All angular variables can be written explicitly as a function of the shower direction $\mathbf{k}$:
\begin{align}\label{eq:pos}
    \omega_{i} = \arccos(\mathbf{k}\cdot\mathbf{\hat{x}}_{i}),\quad l_{i} = \norm{\mathbf{x}_{i}}, \quad \eta_{i} = \operatorname{arctan2}(y_{i}^{\rm sp}, x_{i}^{\rm sp}), \quad \alpha = \arccos(\mathbf{k}\cdot\mathbf{\hat{B}}),
\end{align}
where $\mathbf{x}_{i} = (x_{i}, y_{i}, z_{i})$ is the position of the $i$-th antenna with respect to the emission source, $\mathbf{\hat{x}}_i=\mathbf{x}_{i}/\norm{\mathbf{x}_{i}}$, and $x_{i}^{\rm sp}, y_{i}^{\rm sp}$ are its Cartesian coordinates in the shower-plane frame $\{\mathbf{k} \times \mathbf{B},\ \mathbf{k} \times (\mathbf{k} \times \mathbf{B})\}$, and $\mathbf{\hat{B}}$ is the unit vector of the magnetic field.

\subsection{An equivalent, directly interpretable width parameter}\label{app:reformulation}
The width parameter $\delta \omega$ of Eq.~\eqref{eq:cher} has no direct geometrical meaning. We show here that it can be traded for the width $\Delta \omega$ of the Cherenkov ring at mid-height, which is easier to interpret and to compare between experiments.

\begin{figure}[!ht]
    \centering
    \includegraphics[width=0.7\linewidth]{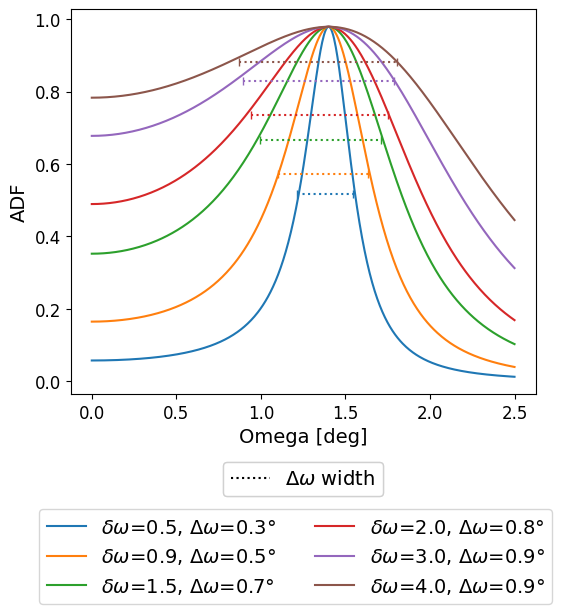}
    \caption{Visualization of $\Delta\omega$, here for $\omega_c = 1.4\degree$. The relation of Eq.~\eqref{eq:deltaomega} holds as long as $\delta \omega\lesssim 2$, i.e. $\Delta\omega \lesssim \omega_c/\sqrt3$.}
    \label{fig:DeltaOmega}
\end{figure}

In the small-angle limit $\tan \omega \simeq \omega$, Eq.~\eqref{eq:cher} reads
\begin{align}\label{eq:cher_smallangle}
    f^{\rm Cherenkov}(\omega) \simeq \frac{1}{1+4\left(\dfrac{\omega^{2}-\omega_{c}^{2}}{\omega_{c}^2\,\delta\omega}\right)^{2}},
\end{align}
with the limiting values
\begin{align}
f^{\rm Cherenkov}(\omega_{c}) = 1, \qquad f^{\rm Cherenkov}(0)=\frac{\delta\omega^{2}}{\delta\omega^{2}+4}.
\end{align}
Let $\beta \in \left[1,\ 1/f(0)\right]$ and let $X$ be such that $f(X) = 1/\beta$. Eq.~\eqref{eq:cher_smallangle} gives
\begin{align}
    1+4\left(\frac{X^{2}/\omega_c^2-1}{\delta\omega}\right)^{2}=\beta
    \quad \logeq \quad
    X^{2}=\omega_c^2\left(1\pm\frac{\delta\omega}{2}\,\sqrt{\beta-1}\right),
\end{align}
so that, taking $X \geq 0$, the function crosses the level $1/\beta$ at
\begin{align}\label{eq:polynom_solution}
X_{1}^{2}=\omega_{c}^{2}\left(1-\frac{\delta\omega}{2}\sqrt{\beta-1}\right),\qquad
X_{2}^{2}=\omega_{c}^{2}\left(1+\frac{\delta\omega}{2}\sqrt{\beta-1}\right).
\end{align}
Writing $X_{1}=\omega_{c}-\Delta\omega_1/2$ and $X_{2}=\omega_{c}+\Delta\omega_2/2$, and defining the mean width $\Delta \omega = (\Delta \omega_1+\Delta \omega_2)/2$, we have
\begin{align}
X_2^{2} - X_1^2=\left(\frac{\Delta\omega_1 + \Delta\omega_2}{2}\right) \left( 2\omega_c + \frac{\Delta\omega_2 - \Delta\omega_1}{2} \right) \simeq 2\,\omega_c \Delta\omega,
\end{align}
where the last step uses $|\Delta\omega_2 - \Delta\omega_1| \ll \omega_c$, valid close to the peak. Combining with Eq.~\eqref{eq:polynom_solution}, $X_2^{2} - X_1^2=\omega_{c}^{2}\,\delta\omega\sqrt{\beta-1}$, gives
\begin{align}
\Delta\omega=\frac{\omega_{c}\,\delta\omega\sqrt{\beta-1}}{2}.
\end{align}
Choosing $\beta$ at mid-height between $f(0)$ and $f(\omega_c)$,
\begin{align}
\frac{1}{\beta} = \frac{f(0) + f(\omega_c)}{2} = \frac{\delta\omega^{2} + 2 }{\delta\omega^{2}+4}
\quad \logeq \quad
\beta -1 = \frac{2}{\delta\omega^{2} + 2},
\end{align}
and therefore
\begin{align}\label{eq:deltaomega}
\Delta \omega = \frac{\omega_c}{\sqrt{2+4/\delta\omega^2}},
\qquad \text{or equivalently} \qquad
\frac{4}{\delta\omega^2} = \left( \frac{\omega_c}{\Delta\omega} \right)^2 - 2 .
\end{align}
The ADF can then be written directly in terms of the ring width:
\begin{align}
    f(\omega)=\frac{1}{1+ \left(\left( \frac{\omega_c}{\Delta\omega} \right)^2 - 2\right)\left(\frac{\tan^2{\omega}}{\tan^2{\omega_c}}-1\right)^{2}}.
\end{align}
Fig.~\ref{fig:DeltaOmega} shows that $\Delta \omega$ as given by Eq.~\eqref{eq:deltaomega} does reproduce the observed width, as long as $\delta \omega\lesssim 2$, i.e. $\Delta\omega \lesssim \omega_c/\sqrt3$. For the value $\delta \omega = 1.2$ adopted in Sec.~\ref{sec:adftuning} and a typical $\omega_c = 1.4\degree$, this gives $\Delta \omega \approx 0.6\degree$.

\section{Optimality of the ADF weights}\label{app:snr_optimality}
We show here that the weights of Eq.~\eqref{eq_heron:weights} maximize the SNR of the coherently summed signal, and that this maximum is reached at the true source position and direction.

From Eq.~\eqref{eq:signal}, the signal on antenna $i$ is $\rbm{S}_i(t) = \mathcal{A}_i\,\rbm{s}_i(t-\Delta t_i(\rbm{X_e}))$, 
with $\mathcal{A}_i$ the ADF amplitude predicted at antenna $i$ from $\rbm{X_e}$ for the true angles $\theta_0$ and $\phi_0$, and $\rbm s_i$ a dimensionless waveform peaking at $1.0$ for $t=0$~ns. For a weight function $w_i$, a source point $\rbm{X_s}=\rbm{X_e}$ and a direction $(\theta,\, \phi)$, the summed signal is
\begin{align}
    \rbm S \left( \theta,\, \phi,\, \rbm{X_s};\ t \right) &= \sum_i w_i(\theta,\, \phi,\, \rbm{X_s}) \rbm{S}_i\left( t + \Delta t_i (\rbm{X_s}) \right) \nonumber \\
    \iff \rbm S \left( \theta,\, \phi,\, \rbm{X_s};\ t \right) &= \sum_i w_i(\theta,\, \phi,\, \rbm{X_s}) \mathcal{A}_i\,\rbm{s}_i(t).
\end{align}
Its maximum amplitude is reached when the peaks of all the $\rbm{s_i}$ add coherently, i.e. at $t=0$~ns where $\rbm{s_i}(0) = 1$ for all $i$, possible only if $\rbm{X_s}=\rbm{X_e}$ (coherence condition), and assuming that the polarization varies slowly across the footprint (i.e., all $\rbm{s_i}$ are parallel\footnote{In reality this is not true for all event configurations, but the misalignment will only change the maximum of the summed signal, not its location.}). Denoting it by $\mathcal{M}$,
\begin{align}
    \mathcal{M}(\theta,\, \phi,\, \rbm{X_s}) = \sum_i w_i(\theta,\, \phi,\, \rbm{X_s}) \mathcal{A}_i \ ,
\end{align}
the Cauchy--Schwarz inequality gives
\begin{align} \label{eq:cauchy_schwarz}
    \frac{\mathcal{M}(\theta,\, \phi,\, \rbm{X_s})}{\|w\|} \leq \| \mathcal{A}\|,
\end{align}
with $\Vert \mathcal{A}\Vert$ the norm of the vector of the $\mathcal{A}_i$, i.e. of the signal maximum amplitudes over all antennas. The inequality becomes an equality if and only if there exists $\gamma>0$ such that $w_i = \gamma \, \mathcal{A}_i$ for all $i$.

Choosing, for any $\rbm{X_s}$, $\theta$ and $\phi$, the weights
\begin{align} \label{eq:weights_demo}
    w_i = f^{\rm ADF}\left( \rbm{p}_i, \rbm{X_s}, \theta, \phi, A=1 \right)\big/\left\| f^{\rm ADF}\left( \rbm{P}, \rbm{X_s}, \theta, \phi, A=1 \right)\right\|,
\end{align}
with $\rbm{P}$ the vector of all antenna positions, we have $\norm{w}=1$ by construction, so that the left-hand side of Eq.~\eqref{eq:cauchy_schwarz} is simply $\mathcal{M}$. The bound is therefore saturated if and only if
\begin{align}\label{eq:equality_condition}
    w_i(\theta,\, \phi,\, \rbm{X_s}) = \frac{\mathcal{A}_i}{\norm{\mathcal{A}}} \qquad \text{for all } i .
\end{align}
The true amplitudes are themselves given by the ADF evaluated at the true parameters, $\mathcal{A}_i = A \, f^{\rm ADF}\qty( \rbm{p}_i, \rbm{X_e}, \theta_0, \phi_0, A=1)$, so the unknown scaling factor $A$ cancels in the ratio $\mathcal{A}_i/\norm{\mathcal{A}}$ and Eq.~\eqref{eq:equality_condition} reduces to the equality of two normalized ADF patterns,
\begin{align}\label{eq:pattern_equality}
    \frac{f^{\rm ADF}\qty( \rbm{p}_i, \rbm{X_s}, \theta, \phi, A=1)}{\norm{f^{\rm ADF}\qty( \rbm{P}, \rbm{X_s}, \theta, \phi, A=1)}}
    = \frac{f^{\rm ADF}\qty( \rbm{p}_i, \rbm{X_e}, \theta_0, \phi_0, A=1)}{\norm{f^{\rm ADF}\qty( \rbm{P}, \rbm{X_e}, \theta_0, \phi_0, A=1)}}
    \qquad \forall\, i .
\end{align}
Assuming the normalized ADF pattern is injective in these parameters, Eq.~\eqref{eq:pattern_equality} holds if and only if $\rbm{X_s} = \rbm{X_e}$, $\theta=\theta_0$ and $\phi=\phi_0$: the metric is maximal at the true source position and direction.\footnote{Injectivity is a property of the full pattern, not necessarily of the pattern sampled by a finite array covering only part of the footprint. Fig.~\ref{fig_heron:interf_adf_map_noise_outlier} shows an event for which two well-separated parameter sets produce sampled patterns that are compatible with the observation.} The cancellation of $A$ in Eq.~\eqref{eq:pattern_equality} is also the reason why the method requires no amplitude calibration (see Sec.~\ref{sec:weights}).

In the presence of Gaussian noise of standard deviation $\sigma$ on each antenna (encompassing the Galactic background and the electronic noise) the sum becomes
\begin{align}
    \rbm S \left( \theta,\, \phi,\, \rbm{X_s};\ t \right) = \sum_i w_i(\theta, \, \phi,\, \rbm{X_s}) \mathcal{A}_i\,\rbm{s}_i(t) + \rbm{z},
\end{align}
with $\rbm{z}$ a Gaussian noise of standard deviation $\sigma \|w\| = \sigma$, since the weights are normalized. The SNR of the summed signal is then
\begin{align} \label{eq:SNR}
    \mathrm{SNR}(\theta,\, \phi,\, \rbm{X_s}) = \frac{\mathcal{M}(\theta,\, \phi,\, \rbm{X_s})}{\sigma} = \frac{\sum_i w_i(\theta,\, \phi,\, \rbm{X_s}) \mathcal{A}_i}{\sigma},
\end{align}
which is again maximal at the true source position and direction. 
The quantity $\mathcal{M}$ above is built from the noise-free amplitudes $\mathcal{A}_i$; it is therefore the true SNR of the coherent signal, in the sense of Sec.~\ref{sec:simu}. What is measurable on data is the corresponding observed SNR, obtained by applying Eq.~\eqref{eq:SNR} to the noisy waveforms, and which is bounded from below since the maximum of the noise is never zero. The two coincide, and maximizing the observed quantity therefore recovers the true source position and direction, only when
\begin{align}\label{eq_heron:recons_validity_condition}
    \mathcal{M} \gg \sigma,
\end{align}
i.e. when the coherent signal dominates the noise pedestal. This condition is verified numerically in Fig.~\ref{fig_heron:scatter_angular_snr}.

\bibliography{biblio}

\end{document}